\documentclass[fleqn,12pt]{article}
\usepackage[numbers,sort&compress]{natbib}
\usepackage{graphics}
\usepackage{graphicx}
\usepackage{amsmath}
\usepackage{amssymb}
\usepackage{pifont}
\usepackage{amsfonts}
\usepackage{caption2}
\usepackage{float}
\usepackage[a4paper,height=25cm,hmargin={2cm,2cm}]{geometry}
\makeatletter
\newcommand\figcaption{\def\@captype{figure}\caption}
\newcommand\tabcaption{\def\@captype{table}\caption}
\makeatother

\begin{document}
\title{Soliton Trapping in Disordered Lattice}
\date{}
\vspace{5mm}
\author{Zhi-Yuan Sun\thanks{
with e-mail address as  sunzhiyuan137@aliyun.com}~, ~Shmuel
Fishman\thanks{ with e-mail address as
fishman@physics.technion.ac.il }
\\{\em Department of Physics,}\\
{\em  Technion-Israel Institute of Technology, Haifa 32000,
Israel}\\
\vspace{2mm}\\ Avy Soffer\thanks{ with e-mail address as
soffer@math.rutgers.edu }
\\{\em Department of Mathematics,}\\
{\em  Rutgers University, New Jersey 08854, U.S.A}\\} \maketitle

\vspace{2cm}
\begin{abstract}
Dynamics of solitons of the Ablowitz-Ladik model in the presence of
a random potential is studied. In absence of the random potential it
is an integrable model and the solitons are stable. As a result of
the random potential this stability is destroyed. In some regime,
for short times particle-like dynamics with constant mass is found.
There is another regime, where particle-like dynamics with varying
mass takes place. In particular an effective potential is found. It
predicts correctly changes in the direction of motion of the
soliton. This potential is a scaling function of time and strength
of the potential, leading to a relation between the first time when
the soliton changes direction and the strength of the random
potential.

\vspace{10mm}

\noindent\emph{PACS number(s)}: 05.45.Yv, 42.65.Tg

\vspace{5mm}\noindent\emph{Keywords}: Ablowitz-Ladik model; random
potential; soliton trapping; particle approach

\end{abstract}

\newpage
\noindent\textbf{I. INTRODUCTION}
\\\hspace*{\parindent}

Solitons are one of the most remarkable manifestations of
nonlinearity. They are found for continuous systems for the
Nonlinear Schr\"{o}dinger Equation (NLSE) in one dimension (1D)
\cite{Sulem}. On the lattice mobile solitons are found for the model
introduced by Ablowitz and Ladik (AL) \cite{Ablowitz}, while for the
ordinary NLSE on a lattice, a mobile soliton is only an approximate
concept. Both lattice and in the continuum disorder tend to affect
and typically destroy solitons. In the present work this will be
studied for a 1D lattice in the framework of the AL model.

For the continuous system, early numerical work of Bronski
\cite{Bronski} indicates that, a NLSE soliton becomes trapped in the
random media when its kinetic energy decreases sufficiently and is
of comparable size to the background potential. Akkermans et.al.
\cite{Akkermans} numerically show that a soliton bounces back and
forth between high potential barriers in the attractive
Bose-Einstein condensates in the framework of the Gross-Pitaevskii
Equation with strong disorder. In addition, like the Anderson
localization of linear waves in random media, some authors relate
the localization of solitons in disordered environment to
\textit{Anderson localization} \cite{Kartashov,Sacha}.

In absence of a random potential the AL model is
\begin{equation}
i \frac{\partial \psi_n}{\partial t} = -(\psi_{n-1} + \psi_{n+1}) (1
+ |\psi_n|^2)~, \label{1.1}
\end{equation}
where $\psi_n$ is the wavefunction on site $n$ at time $t$. The
integrability is manifested by the existence of a mobile soliton
solution \cite{Cai,Kevrekidis0}
\begin{equation}
\psi_n(t) = \frac{\sinh(\mu)}{\cosh[\mu(n-x)]} \exp[i k (n-x) + i
\alpha]~, \label{2}
\end{equation}
where the time-dependent parameters $x$ and $\alpha$ can be
expressed as
\begin{subequations}
\begin{align}
&\hspace{0mm} \dot{x} = 2 \frac{\sinh(\mu)}{\mu} \sin(k)~,\label{2.1}\\
&\hspace{0mm} \dot{\alpha} = 2 [\cosh(\mu)\cos(k) + \frac{k}{\mu}
\sinh(\mu)\sin(k)]~.\label{2.2}
\end{align}
\end{subequations}
From (\ref{2}) we see that $\frac{1}{\mu}$ characterizes the width
of the soliton and $x$ is its center. On the other hand, the AL
equation has two conserved quantities, the first of which can be
defined as the mass of the soliton solution
\cite{Cai,Kevrekidis0,Kevrekidis}
\begin{equation}
M_s = \sum_{n=-\infty}^{\infty} \ln(1+|\psi_n|^2)~, \label{3}
\end{equation}
while the second can be defined as the momentum of the motion
\cite{Cai,Kevrekidis0,Kevrekidis}
\begin{equation}
P = i \sum_{n=-\infty}^{\infty} (\psi_n \psi_{n+1}^* - \psi_n^*
\psi_{n+1})~,\label{4}
\end{equation}
where $*$ denotes the complex conjugation. For the soliton solution
(\ref{2}) we can calculate that (also can see Appendix of
\cite{Cai})
\begin{equation}
M_s = 2\mu~,\label{4.1}
\end{equation}
and
\begin{equation}
P = M_s \dot{x} = 4 \sinh(\mu) \sin(k)~.\label{4.2}
\end{equation}
Therefore $M_s$ can be indeed considered as the mass of the soliton.

In the present work we will study solitons for the AL model with a
random potential defined by
\begin{equation}
i \frac{\partial \psi_n}{\partial t} = -(\psi_{n-1} + \psi_{n+1}) (1
+ |\psi_n|^2) + \varepsilon_n \psi_n~, \label{1}
\end{equation}
where $\varepsilon_n$ are independent random variables uniformly
distributed in the interval $[-\frac{W}{2}, \frac{W}{2}]$. For the
continuous version, two early reviews \cite{Bass,Gredeskul} have
addressed the propagation of solitons in disordered systems; in the
works by Bronski \cite{Bronski1,Bronski} and Garnier
\cite{Garnier1}, they show two regimes for the NLSE soliton
propagation. In one regime, the soliton mass decays while its
velocity approaches a constant; in the other regime, the soliton
mass approaches a constant while its velocity decays very slowly.
Garnier further applied a perturbation theory of the inverse
scattering transform to confirm that similar two regimes are found
for the AL solitons with on-site random potential (in the limit of
zero randomness) \cite{Garnier2}. Which regime is relevant depends
on the value of the initial mass of the AL soliton. For large $\mu$,
the mass approaches a constant, while for small $\mu$, the velocity
approaches a constant.

However, we will show numerically, for the weak randomness, the
large soliton will be trapped before its velocity decreases to zero.
Additionally, we will find a regime in which the AL soliton has
possibility to be accelerated on the average by the randomness. We
will also characterize the regime where soliton can be trapped by
the disorder using a particle approach.
\\

\noindent\textbf{II. SOLITON PROPAGATION IN A DISORDERED LATTICE}
\\\hspace*{\parindent}

In this section we will study the solution of Eq.~(\ref{1})
numerically and semianalytically in order to develop an intuitive
picture of the soliton dynamics. The initial soliton is the one
found for a chain without disorder, given by (\ref{2}) with
\begin{subequations}\label{4.3}
\begin{align}
&\hspace{0mm} x=0, ~~~\dot{x}>0~,\label{4.3a}\\
&\hspace{0mm} M_s(t=0)=2\mu~.\label{4.3b}
\end{align}
\end{subequations}
The numerical solution is obtained propagating the soliton by
Eq.~(\ref{1}). To save computer resources we use a coordinate system
moving with the center of mass of the soliton, consisting of $N$
sites, centered on soliton. The computation is performed using a
4th-order Runge-Kutta type algorithm in time, and an absorbing-wave
boundary condition on sites. We define the following quantities:

soliton mass
\begin{equation}
M_s^{(N)}=\sum_n^{N} \ln(1+|\psi_n|^2)~;\label{4.4}
\end{equation}
center of mass coordinate
\begin{equation}
x^{(N)}=\sum_n^{N} n\ln(1+|\psi_n|^2)/M_s^{(N)}~,\label{4.5}
\end{equation}
and the second moment
\begin{equation}
m_2^{(N)}=\sum_n^{N} (n-x^{(N)})^2 \ln(1+|\psi_n|^2)~,\label{4.6}
\end{equation}
while the soliton velocity is
\begin{equation}
v=\Delta x^{(N)}/\Delta t~,\label{4.7}
\end{equation}
where $\Delta x^{(N)}$ is the change of $x^{(N)}$ during the time
interval $\Delta t$ (here we use $\Delta t=0.001$). In addition, two
parameters, that characterize the soliton are introduced in
simulation: one is the amplitude of the soliton
\begin{equation}
A_s = \max_n |\psi_n|^2~,\label{4.8}
\end{equation}
the other one is the soliton width, defined as the minimum $N_w$
satisfying
\begin{equation}
\sum_{-(N_w-1)/{2}}^{(N_w-1)/{2}} \ln(1+|\psi_n|^2)/M_s^{(N)}
\geq1-\delta~,\label{4.9}
\end{equation}
here $\delta = 0.01$ [note that, in simulation we first find the
peak position of the soliton, then (\ref{4.9}) is calculated with
this position as the center].

There are basically 3 regimes characterized by the initial value of
$\mu$:
\begin{subequations}\label{4.91}
\begin{align}
&\hspace{0mm} (A)~~~ \mu\gg1~;\\
&\hspace{0mm} (B)~~~ \mu\approx1~;\\
&\hspace{0mm} (C)~~~ \mu\ll1~.
\end{align}
\end{subequations}

\noindent\textbf{A. The regime $\mu\gg1$}
\\\hspace*{\parindent}

We choose an AL soliton with $\mu=3$, which has more than $99\%$
mass concentrating in 3 lattice sites ($N_w=3$). The reason for
picking up a soliton of such narrow width is based on the fact that
it is compact enough to admit low level of mass radiation resulting
of randomness. Such low-level radiation is necessary for observing
possible soliton acceleration in our numerical simulation. The
initial velocity of the soliton is chosen as $\dot{x}(t=0)=1$, and
one realization of the random potential with $W=0.1$ is used (note
that $0.04\lesssim W \lesssim0.1$ can be seen as the weak randomness
in our discussion, below this interval is considered as the very
weak randomness where the soliton dynamics may approach the one in
the limit of zero randomness as in \cite{Garnier2}).

Assuming the random potential is a perturbation, the approximate
equations for the various parameters in this potential can be
derived following the work of Cai et.al. \cite{Cai}. The resulting
equations derived in Appendix~A are
\begin{subequations}\label{5}
\begin{align}
&\hspace{0mm} \dot{\mu} = 0~,\label{5a}\\
&\hspace{0mm} \dot{x} = \frac{2\sinh(\mu)}{\mu} \sin(k)~,\label{5b}\\
&\hspace{0mm} \dot{k} = \sinh^2(\mu) \sum^{+\infty}_{n=-\infty}
\frac{\varepsilon_n \tanh[\mu (n-x)]}{\cosh[\mu (n+1-x)] \cosh[\mu
(n-1-x)]}~, \label{5c}\\
&\hspace{0mm} \dot{\alpha} = 2 [\cosh(\mu)\cos(k) + \frac{k}{\mu}
\sinh(\mu)\sin(k)] \notag \\
&\hspace{7mm} + \sinh^2(\mu) \sum^{+\infty}_{n=-\infty}
\frac{\varepsilon_n (n-x) \tanh[\mu (n-x)]}{\cosh[\mu (n+1-x)]
\cosh[\mu (n-1-x)]} \notag \\
&\hspace{7mm} - \sinh(\mu)\cosh(\mu) \sum^{+\infty}_{n=-\infty}
\frac{\varepsilon_n }{\cosh[\mu (n+1-x)] \cosh[\mu
(n-1-x)]}~.\label{5d}
\end{align}
\end{subequations}
Eqs.~(\ref{5}) are integrated numerically with the initial
conditions (\ref{4.3}). The algorithm used is the 4th-order
Runge-Kutta with $\Delta t=0.001$, and the summations are truncated
to a finite window around the center of mass of the soliton. The
values of the parameters $\mu$, $k$, $x$, and $\alpha$ are inserted
in (\ref{2}). We refer to this solution as the semianalytical
solution. We compare this solution with the numerical integration of
Eq.~(\ref{1}) (referred as the numerical solution), and the results
are presented in Fig.~\ref{A}. In Fig.~\ref{A}(a) we compare the
center of mass coordinate $x$ of (\ref{2}) and (\ref{4.5}) found in
the semianalytical calculation with the numerical solution, and
small acceleration is presented for this type of soliton [see also
Fig.~\ref{A}(f)]. It is found that on the average the velocity of
the semianalytical solution is somewhat larger than the one found
numerically. In Fig.~\ref{A}(b) the velocity of the center of mass
and the second moment are presented. The plots for the velocity are
zoomed in Figs.~\ref{A}(c)-(e). We note that for $t<100$ there is
excellent agreement between the numerical and the semianalytical
results. At the time $t>100$ the second moment increases rapidly,
therefore the approximation (\ref{5}) is not justified anymore, and
large deviation between the two solutions are shown. In
Fig.~\ref{A}(f),
the semianalytical velocity for longer time and its linear fit are presented.\\

\noindent\textbf{B. The regime $\mu\approx1$}
\\\hspace*{\parindent}

As a representative example in this regime, we study a soliton with
$\mu=1$, moving in one realization of the random potential with
$W=0.1$. We solve numerically Eq.~(\ref{1}) with the initial
condition (\ref{4.3}). In this case the initial width of the soliton
is 7 sites ($N_w=7$). The results are presented in Fig.~\ref{B}. We
find that the center of mass $x$ moves monotonically to the right
till a time $t=T_c$ when oscillations start [see Fig.~\ref{B}(a)].
From Fig.~\ref{B}(b) we see that the velocity decreases
monotonically for $t<T_c$ and oscillates for $t>T_c$. The period of
these oscillations decreases with time. From Fig.~\ref{B}(c) we see
that the mass decreases with time and in the first stage this
decrease is rapid, therefore the approximation (\ref{5}) fails.
Finally the mass approaches a nonvanishing constant. The interesting
phenomenon we find is the trapping of the soliton for $t>T_c$ as a
result of randomness. The particle aspect of this dynamics will be
discussed
in the next section.\\

\noindent\textbf{C. The regime $\mu\ll1$}
\\\hspace*{\parindent}

In this regime, the soliton has a larger width, and it is easier to
lose its mass through radiation. With a limit of zero randomness,
Garnier \cite{Garnier2} shows that the soliton propagates with its
mass decreasing to zero, and its velocity to a nonvanishing
constant. In fact, the radiation induces a remarkable deformation on
the soliton profile after some time of propagation if the randomness
is not weak enough. In Fig.~\ref{C} we present an example with
$\mu=0.5$ [where the initial width of the soliton is 11 sites
($N_w=11$)] and $W=0.1$. The initial condition is (\ref{4.3}). From
Fig.~\ref{C}(a) we see that the soliton spreads, and radiates its
mass over $300$ sites in the time $t=1000$. From Figs.~\ref{C}(b)
and (c) we conclude that the mass and velocity decrease. Such
decrease of velocity contains some short time intervals where the
velocity oscillates approximately near a constant. The similar time
intervals have been observed in the work of Franzosi et.al.
\cite{Franzosi}, as they studied the mobile discrete breathers
propagating on very weak backgrounds in the framework of discrete
NLSE. Their time intervals appear to be much longer, with weaker
velocity oscillation, since their background perturbation is very
weak (of the level $10^{-4}\sim10^{-3}$). However, we have not
observed the trapping behavior in this regime,
especially before the soliton undergoes a large deformation.\\

\noindent\textbf{III. PARTICLE APPROACH FOR SOLITON TRAPPING IN DISORDERED AL LATTICE}
\\\hspace*{\parindent}

In this section we will study the question: Can an AL soliton in a
weak random potential be considered as particle?

We will focus on the soliton trapping in the second regime where
$\mu\approx1$, and try to give a particle description of the
trapping behavior. We start from the momentum (\ref{4}), and assume
it still to be the momentum for the model (\ref{1}) when the random
potential is weak. Taking a derivative on both sides of
Eq.~(\ref{4}) with respect to $t$, and substituting Eq.~(\ref{1})
into the result, we can obtain
\begin{equation}
\frac{dP}{dt} = 2 \sum_{n=-\infty}^{+\infty} \textmd{Re} (\psi_n
\psi_{n+1}^*) (\varepsilon_n-\varepsilon_{n+1})~,\label{6}
\end{equation}
where Re means the real part. For derivation see Appendix~B. With
the assumption that the soliton is particle-like, Eq.~(\ref{6}) can
be viewed as the variation rate of its momentum. On the other hand,
$dP/dt$ by (\ref{4.2}) can be also written as
\begin{equation}
\frac{dP}{dt} = \frac{d M_s}{dt}v + M_s \frac{d^2
x}{dt^2}~.\label{7}
\end{equation}
Notice that, due to the mass radiation, the term $dM_s/dt$ in
Eq.~(\ref{7}) can not be neglected, especially before soliton
trapping. Therefore, we can write the randomness-generated
\textit{force} in two ways, one is directly
\begin{equation}
F_1= M_s \frac{d^2 x}{dt^2}~,\label{6.1}
\end{equation}
using (\ref{6}) and (\ref{7}) we derive
\begin{equation}
F_2 = 2 \sum_{n=-\infty}^{+\infty} \textmd{Re} (\psi_n \psi_{n+1}^*)
(\varepsilon_n-\varepsilon_{n+1}) - \frac{d M_s}{dt}v~.\label{6.2}
\end{equation}
The test of the particle-like picture is performed by comparing the
forces $F_1$ and $F_2$ presented in Fig.~\ref{D}. Excellent
agreement is found. These results strongly support the description
of solitons as particles. In this picture with the force $F_2$ we
associate work done on the soliton that decreases its kinetic energy
with an effective potential
\begin{equation}
U(t) = U_0 - \int_{t_0}^t F_2 v dt^{'}~,\label{8}
\end{equation}
where $U_0$ is an parameter which can be viewed as the initial
energy to be determined as the constant that leads the mean of $U$,
over the time interval of trapping in simulations, to be zero, i.e.,
$U_0=<\int_{t>T_c} F_2vdt>$. With the same data, we plot both of
this effective potential $U(t)$ and soliton velocity in
Fig.~\ref{D}(e). It shows that the first reflection
($T_c\approx1100$), with the soliton velocity changing its sign,
occurs at a peak position of the effective potential. Also other
changes in the direction of motion of the soliton take place at
maxima of $U(t)$ as can be seen from Fig.~\ref{D}(e). This is a
direct result of (\ref{8}) since
\begin{equation}
\frac{dU(t)}{dt} = -F_2 v~,\label{8.05}
\end{equation}
therefore $\frac{dU(t)}{dt}=0$ implies either $F_2=0$ or $v=0$. \\

\noindent\textbf{IV. SCALING OF THE TRAPPING TIME $T_c$}
\\\hspace*{\parindent}

In this section we demonstrate that there exists a scaling relation
between the time $T_c$ when the trapping starts and the random
potential strength $W$. In Figs.~\ref{E}(a) and (b) $T_c$ is plotted
as a function of $W$. It is found that
\begin{equation}
T_c \sim W^{-\eta}~,\label{8.1}
\end{equation}
with $\eta=2.32\pm0.41$. For each $W\in[0.06,0.1]$, we average the
function $U(t)$ of (\ref{8}) over 6 different realizations to derive
$U_{ave}(t)$, and plot it till the minimum value of $T_c$ of 6
realizations in Fig.~\ref{E}(c). From Fig.~\ref{E}(d), we see that
$U_{ave}$ is related to $W$ and $t$ via the combination $tW^2$. This
suggests the scaling relation
\begin{equation}
U_{ave} \approx \Gamma(t W^2)~,\label{9}
\end{equation}
where $\Gamma$ is the scaling function. If trapping starts at the
same value of $\Gamma$, one finds
\begin{equation}
T_c \propto W^{-2}~.\label{10}
\end{equation}

Here we want to give some comments on $T_c$. In forming (\ref{8.1})
of this section, we use one realization of random numbers uniformly
distributed in $[-1,1]$, but multiplied by the strength $W/2$, as
the random potential. Since change of $T_c$ is obvious for small
variation of $W$ [see Figs.~\ref{E}(a) and (b)], it can generally
reflect the scaling relation. One may arrange multiple realizations
of random potential to compute the mean value of $T_c$, as well as
its variance possibly, in the statistic sense.

On the other hand, if the randomness is very weak ($W\lesssim0.04$),
Eq.~(\ref{10}) seems to be no longer valid since the soliton can
propagate without reflection for very long time as $U_{ave}$
approaches constant (long-time simulation reveals such feature). In
this regime, before trapping, the soliton with its mass as almost a
constant, loses its velocity very slowly during the very long time
propagation in the random potential. In particular, $T_c$ can reach
to about $10^5$ for $W=0.02$, and much longer for weaker randomness.
We consider this behavior of the soliton to be similar as the
\textit{moving breather} in the discrete NLS
lattice \cite{Franzosi,Neff} under very weak noise. \\

\noindent\textbf{IV. SUMMARY AND CONCLUSIONS }
\\\hspace*{\parindent}

Dynamics of solitons in random potentials was studied in the
framework of the Ablowitz-Ladik model \cite{Ablowitz}. In particular
we explored the question when can a soliton be considered as a
particle and what are the conditions for trapping of solitons in a
random potential. The behavior was classified into three regimes
specified by Eq.~(\ref{4.91}). In the regime $\mu\gg1$ for short
times the approximation (\ref{5}) holds. In particular $\mu$ changes
some time resulting in the change of the soliton width. This
destroys the semianalytic solution resulting of (\ref{5}) as is
clear from Fig.~\ref{A}. An improved approach will be subject of
future studies. For $\mu\ll1$ the soliton spreads very quickly and
the potential picture is not appropriate.

The most interesting regime is when $\mu\approx1$. The most
interesting phenomenon is that the soliton is trapped and moves as a
particle with varying mass. The equality of $F_1=F_2$ that is
demonstrated in Fig.~\ref{D} is a strong evidence for the particle
nature. The velocity changes its direction at some maxima of the
potential (\ref{8}) as can be seen from Fig.~\ref{D}(e) and as
expected from (\ref{8.05}). Better understanding of the potential
$U(t)$ and its relation to the average of the random potential over
the profile of the soliton will be left for future studies. Finally
we found that $T_c$, the first time when the velocity changes
direction, scales with the strength of the random potential
according to (\ref{9}) and the potential is scaling function of time
and the strength of random potential (\ref{8.1}). This may signal
the existence of an underlaying statistical theory that should be
explored in the future.\\

\noindent\textbf{ACKNOWLEDGEMENT}
\\\hspace*{\parindent}
Z.-Y. S. acknowledges the support in part at the Technion by a
fellowship of the Israel Council for Higher Education. This work was
partly supported by the Israel Science Foundation (ISF-1028), by the
US-Israel Binational Science Foundation (BSF-2010132), by the USA
National Science Foundation (NSF DMS 1201394) and by the Shlomo
Kaplansky academic chair.

\newpage
\appendix
\noindent\textbf{Appendix~A }
\\\hspace*{\parindent}

Refs.~\cite{Cai} show that for an AL model with a perturbation term
\begin{equation}
i \frac{\partial \psi_n}{\partial t} = -(\psi_{n-1} + \psi_{n+1}) (1
+ |\psi_n|^2) + R_n~, \label{A1}\tag{A.1}
\end{equation}
the soliton parameters in (\ref{2}) in the adiabatic approximation
satisfy the following evolution equations
\begin{subequations}\label{A2}
\begin{align}
&\hspace{0mm} \dot{\mu} = \sinh(\mu) \sum^{+\infty}_{n=-\infty}
\frac{\cosh[\mu (n-x)] \textrm{Im}(r_n)}{\cosh[\mu (n+1-x)]
\cosh[\mu(n-1-x)]}~,\label{A2a}\tag{A.2a}\\
&\hspace{0mm} \dot{x} = \frac{2\sinh(\mu)}{\mu} \sin(k)
+\frac{\sinh(\mu)}{\mu} \sum^{+\infty}_{n=-\infty}
\frac{(n-x)\cosh[\mu (n-x)] \textrm{Im}(r_n)}{\cosh[\mu (n+1-x)]
\cosh[\mu(n-1-x)]}~,\label{A2b}\tag{A.2b}\\
&\hspace{0mm} \dot{k} = \sinh(\mu) \sum^{+\infty}_{n=-\infty}
\frac{\sinh[\mu (n-x)] \textrm{Re}(r_n)}{\cosh[\mu (n+1-x)]
\cosh[\mu(n-1-x)]}~, \label{A2c}\tag{A.2c}\\
&\hspace{0mm} \dot{\alpha} = 2 [\cosh(\mu)\cos(k) + \frac{k}{\mu}
\sinh(\mu)\sin(k)] \notag \\
&\hspace{7mm} + \sinh(\mu) \sum^{+\infty}_{n=-\infty} \frac{(n-x)
\sinh[\mu (n-x)] \textrm{Re}(r_n)}{\cosh[\mu (n+1-x)]
\cosh[\mu (n-1-x)]} \notag \\
&\hspace{7mm} - \cosh(\mu) \sum^{+\infty}_{n=-\infty}
\frac{\cosh[\mu (n-x)] \textrm{Re}(r_n) }{\cosh[\mu (n+1-x)]
\cosh[\mu (n-1-x)]}\notag\\
&\hspace{7mm} +k\frac{\sinh(\mu)}{\mu} \sum^{+\infty}_{n=-\infty}
\frac{(n-x) \cosh[\mu (n-x)] \textrm{Im}(r_n) }{\cosh[\mu (n+1-x)]
\cosh[\mu (n-1-x)]}~,\label{A2d}\tag{A.2d}
\end{align}
\end{subequations}
where $r_n = R_n \exp[-ik(n-x)-i\alpha]$. For Eq.~(\ref{1}) with the
solution form (\ref{2}) we have
\begin{equation}
r_n = \frac{\varepsilon_n
\sinh(\mu)}{\cosh[\mu(n-x)]}~.\label{A3}\tag{A.3}
\end{equation}
Substituting (A.3) into (A.2) we obtain Eqs.~(\ref{5}). \\

\noindent\textbf{Appendix~B }
\\\hspace*{\parindent}

In this appendix, we will show the derivation of Eq.~(\ref{6}). Take
a derivative on both sides of Eq.~(\ref{4}) with respect to $t$, one
can obtain
\begin{equation}
\frac{dP}{dt} = i \sum_{n=-\infty}^{+\infty} \left(\frac{\partial
\psi_n}{\partial t} \psi_{n+1}^* + \psi_n \frac{\partial
\psi_{n+1}^*}{\partial t} - \frac{\partial \psi_n^*}{\partial t}
\psi_{n+1} - \psi_n^* \frac{\partial \psi_{n+1}}{\partial
t}\right)~.\label{B1}\tag{B.1}
\end{equation}
With Eq.~(\ref{1}), we derive the following sets
\begin{subequations}\label{B2}
\begin{align}
&\hspace{0mm} \frac{\partial \psi_n}{\partial t} =
i[(\psi_{n-1}+\psi_{n+1})(1+\psi_n \psi_n^*) - \varepsilon_n
\psi_n]~,\label{B2a}\tag{B.2a}\\
&\hspace{0mm} \frac{\partial \psi_n^*}{\partial t} =
-i[(\psi_{n-1}^*+\psi_{n+1}^*)(1+\psi_n \psi_n^*) - \varepsilon_n
\psi_n^*]~,\label{B2b}\tag{B.2b}\\
&\hspace{0mm} \frac{\partial \psi_{n+1}}{\partial t} =
i[(\psi_{n}+\psi_{n+2})(1+\psi_{n+1} \psi_{n+1}^*) -
\varepsilon_{n+1} \psi_{n+1}]~,\label{B2c}\tag{B.2c}\\
&\hspace{0mm} \frac{\partial \psi_{n+1}^*}{\partial t} =
-i[(\psi_{n}^*+\psi_{n+2}^*)(1+\psi_{n+1} \psi_{n+1}^*) -
\varepsilon_{n+1} \psi_{n+1}^*]~.\label{B2d}\tag{B.2d}
\end{align}
\end{subequations}
Substituting (B.2) into (B.1), after simplification, we can obtain
\begin{subequations}
\begin{align}
&\hspace{0mm} \frac{dP}{dt} = \sum_{n=-\infty}^{+\infty}
[\varepsilon_n(\psi_n\psi_{n+1}^*+\psi_n^*\psi_{n+1}) -
\varepsilon_{n+1}(\psi_n\psi_{n+1}^*+\psi_n^*\psi_{n+1})]\notag\\
&\hspace{6.5mm} = 2 \sum_{n=-\infty}^{+\infty} \textmd{Re} (\psi_n
\psi_{n+1}^*) (\varepsilon_n-\varepsilon_{n+1})~.\notag
\end{align}
\end{subequations}
Thus Eq.~(\ref{6}) is derived.

\newpage
\noindent\textbf{Figure captions and figures}
\vspace{2cm}
\\[\intextsep]
\begin{minipage}{\textwidth}
\renewcommand{\captionfont}{ }
\renewcommand{\captionlabelfont}{ }
\vspace{-2.6cm}\centering
\includegraphics[scale=0.6]{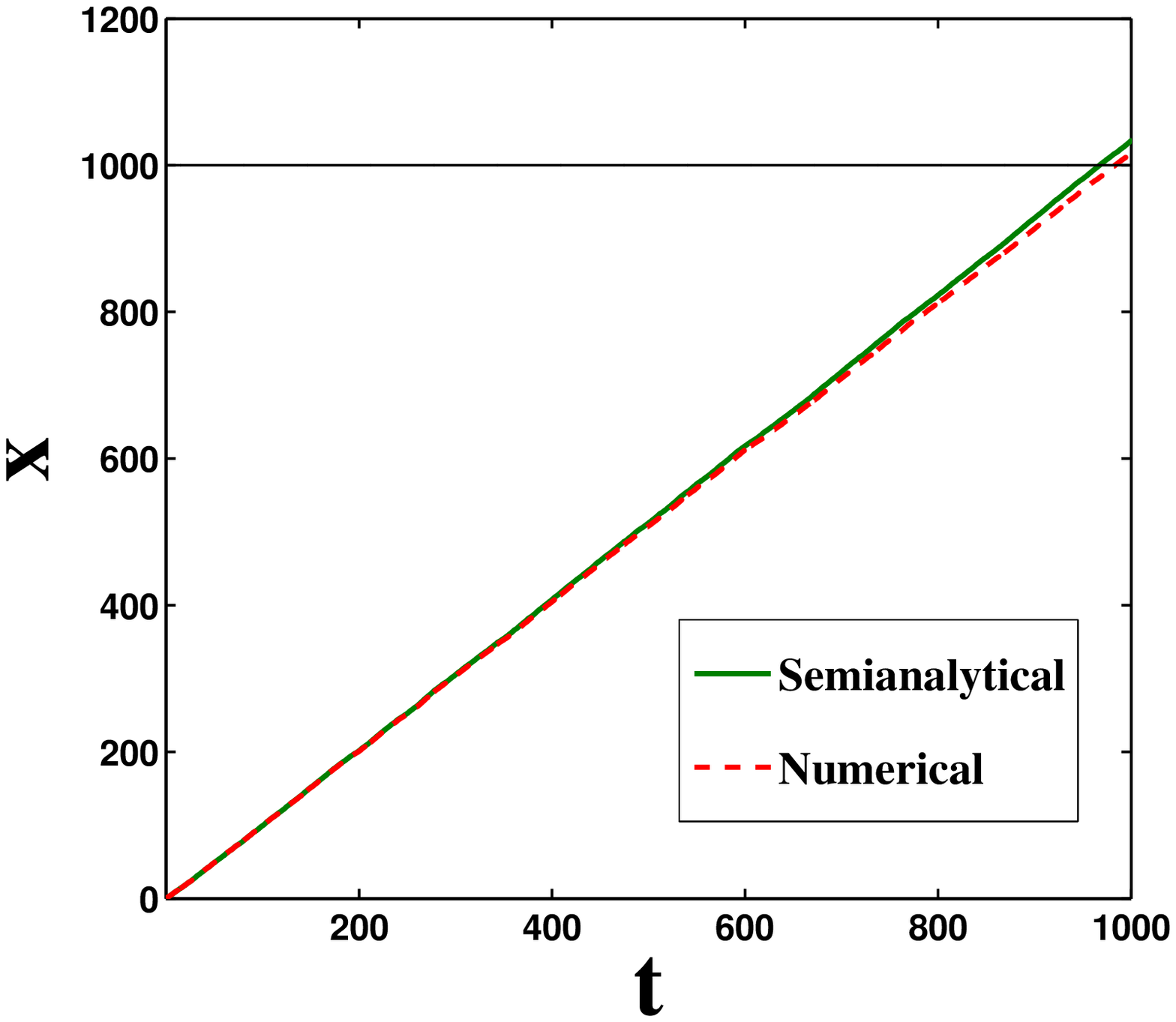}
\vspace{-0.4cm} {\center\footnotesize\hspace{0cm}\textbf{Figure 1(a)}}\\
\vspace{5mm}
\includegraphics[scale=0.7]{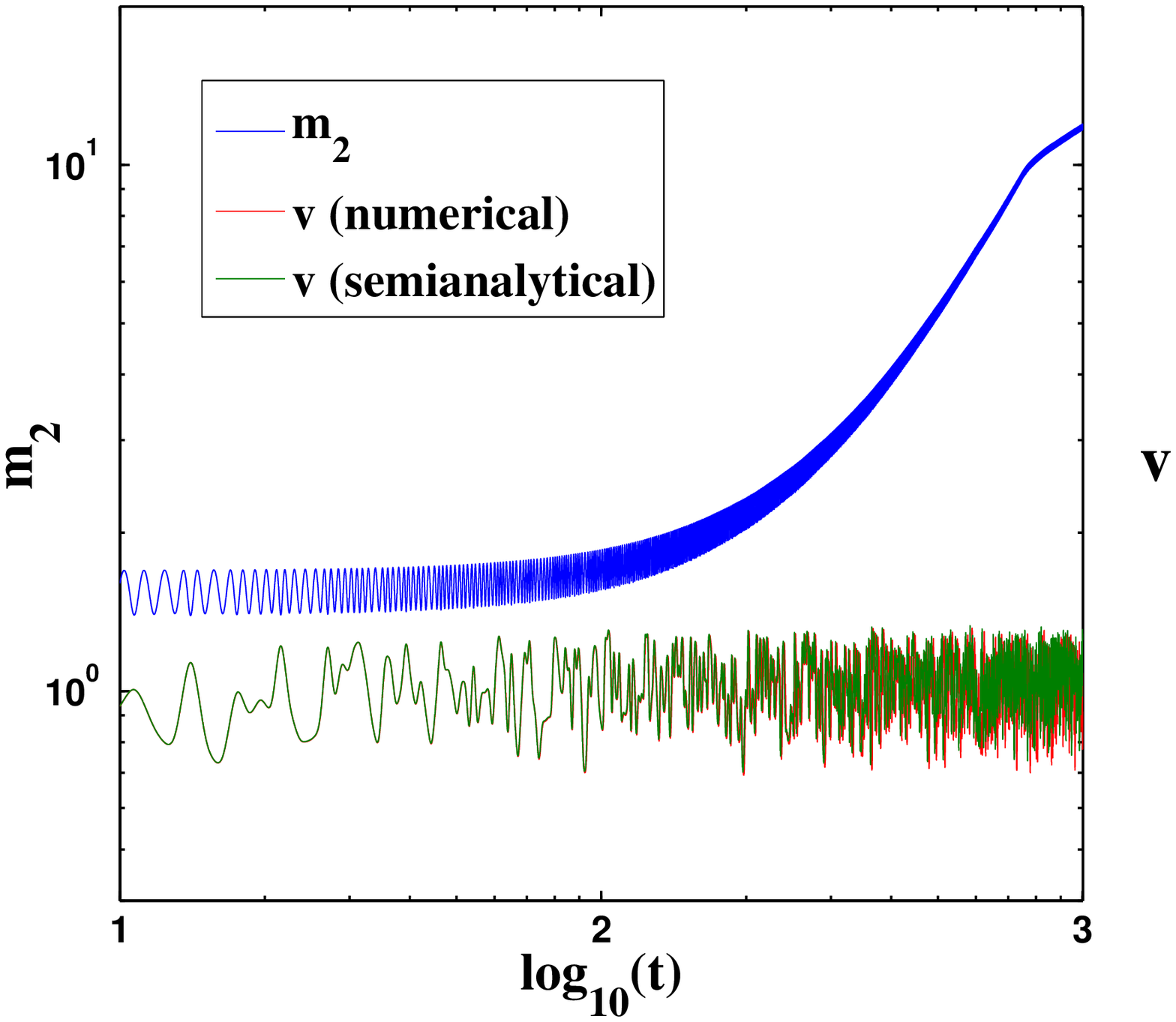}
\vspace{-0.4cm} {\center\footnotesize\hspace{0cm}\textbf{Figure 1(b)}}\\
\notag
\end{minipage}
\\[\intextsep]
\\[\intextsep]
\begin{minipage}{\textwidth}
\renewcommand{\figurename}{FIGURES }
\renewcommand{\captionfont}{ }
\renewcommand{\captionlabelfont}{ }
\vspace{-1.7cm}\centering
\includegraphics[scale=0.43]{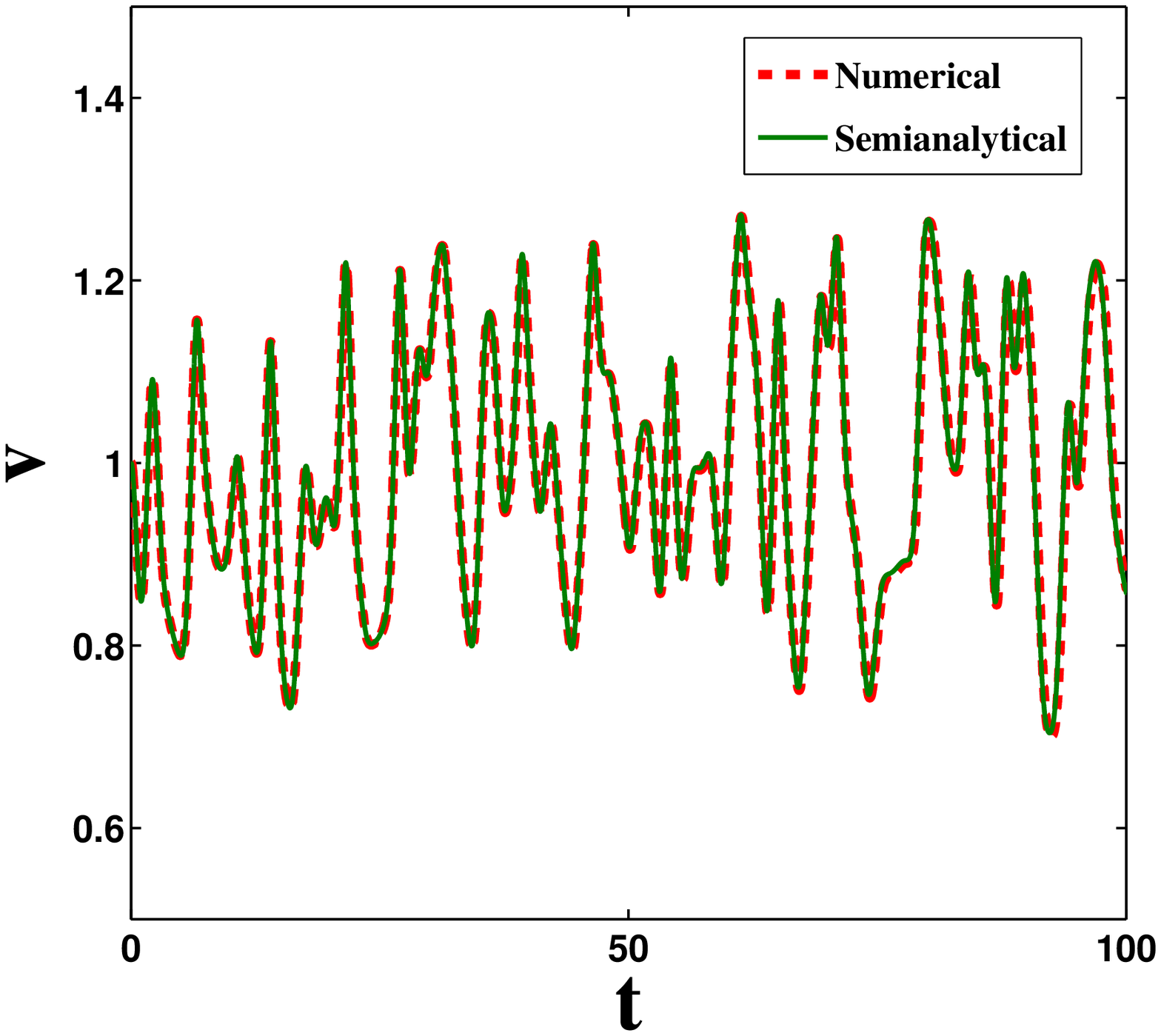}
\includegraphics[scale=0.43]{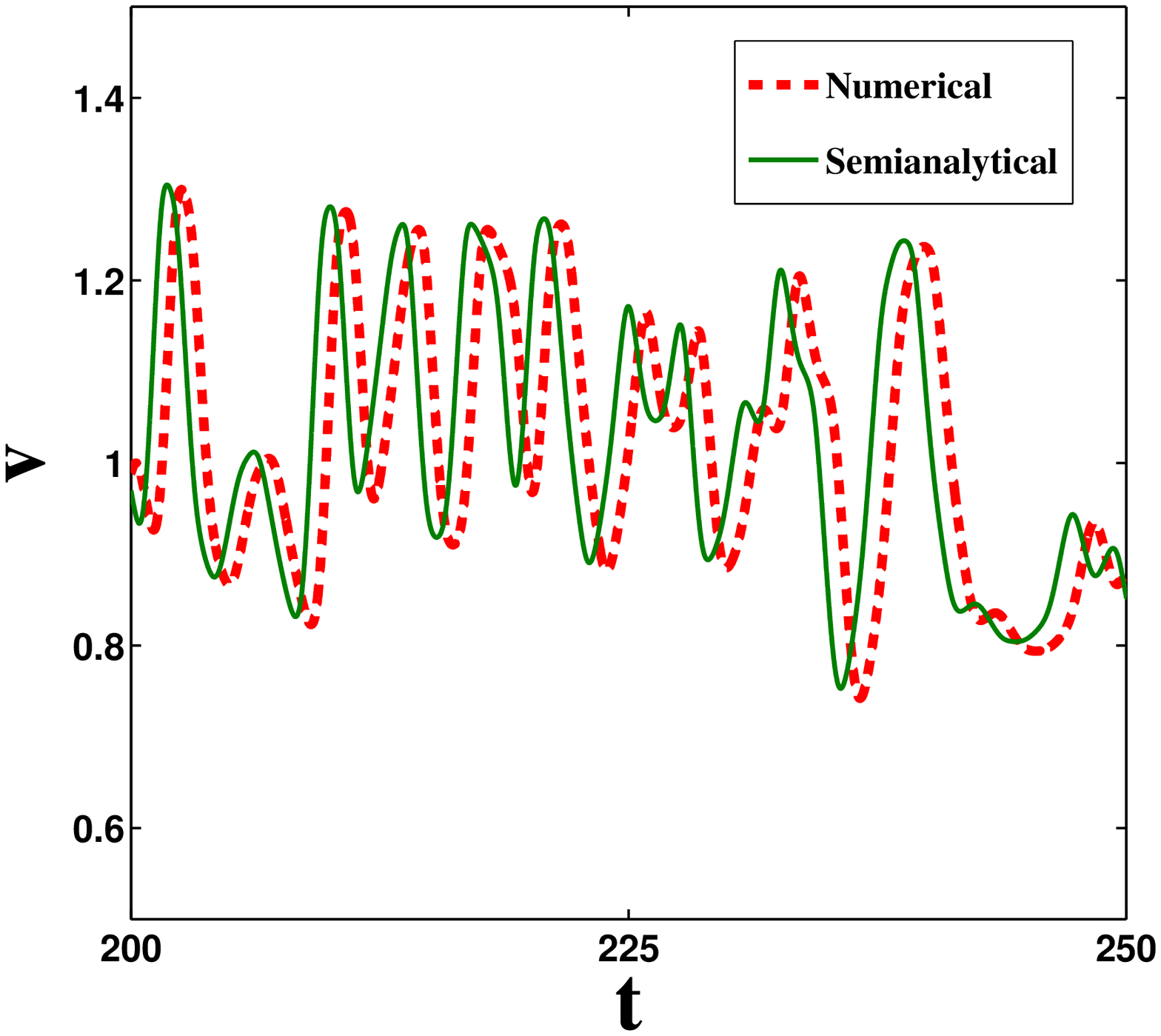}
\vspace{-1.3cm} {\center\footnotesize\hspace{0cm}\textbf{Figure 1(c)}
\hspace{6.6cm}\textbf{Figure 1(d)}}\\
\vspace{2mm}
\includegraphics[scale=0.43]{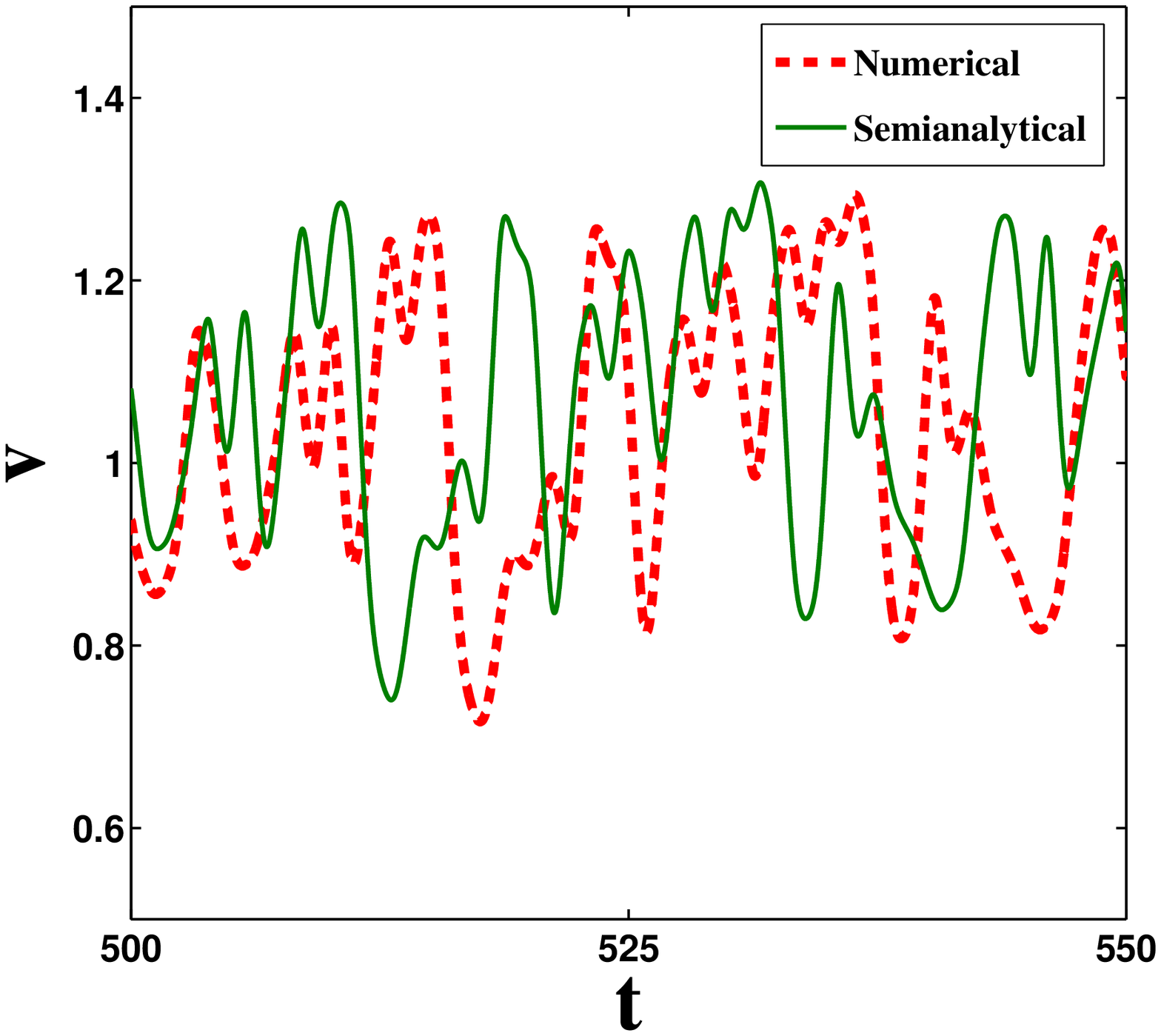}
\vspace{-0.7cm} {\center\footnotesize\hspace{0cm}\textbf{Figure 1(e)}}\\
\vspace{2mm}
\includegraphics[scale=0.45]{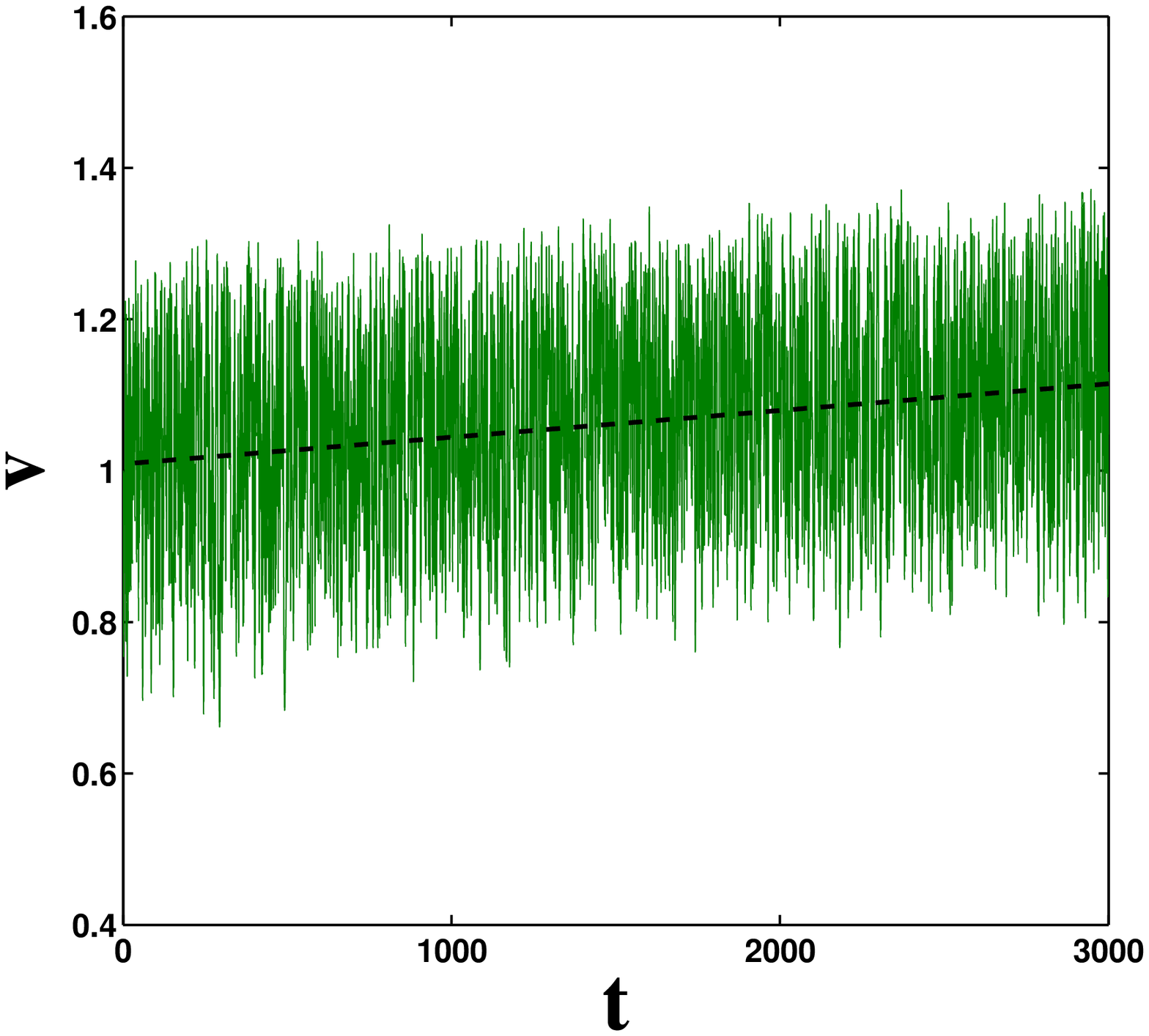}
\vspace{-0.7cm} {\center\footnotesize\hspace{0cm}\textbf{Figure
1(f)}} \figcaption{AL soliton acceleration by one realization of
random potential [$\mu=3$, $v(t=0)=1$, and $W=0.1$]. (a) Comparison
of the center of mass between the semianalytical and numerical
results. (b) Comparison of the semianalytical and numerical
velocities, and the second moment $m_2$. (c)-(e) Zoomed views of the
velocity comparison in three different time intervals. (f)
Semianalytical velocity by solving Eqs.~(\ref{5}), and the dashed
line is the linear fit of the data.   } \label{A}
\end{minipage}
\\[\intextsep]
\\[\intextsep]
\begin{minipage}{\textwidth}
\renewcommand{\figurename}{FIGURES }
\renewcommand{\captionfont}{ }
\renewcommand{\captionlabelfont}{ }
\vspace{-1cm}\centering
\includegraphics[scale=0.42]{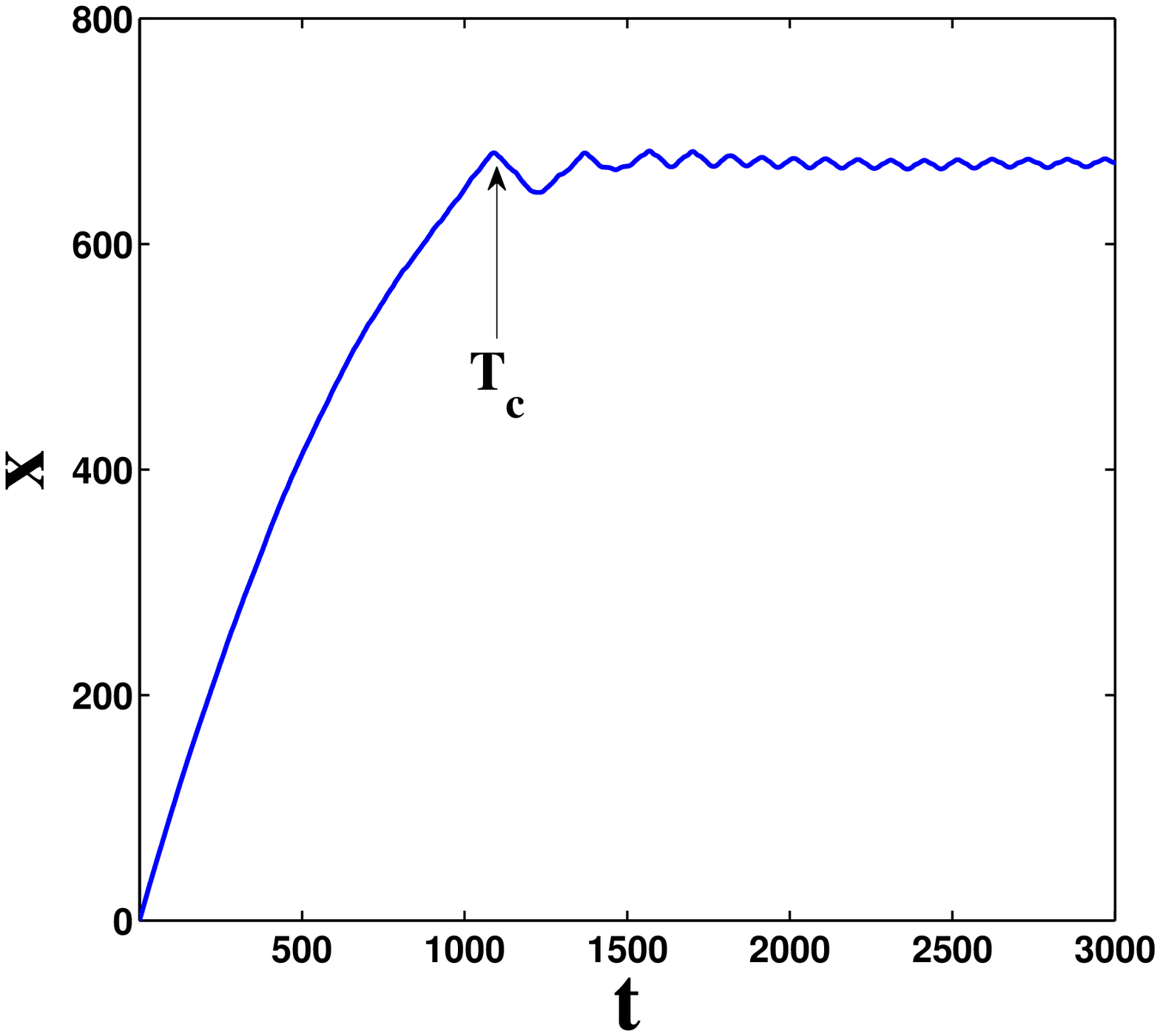}
\includegraphics[scale=0.42]{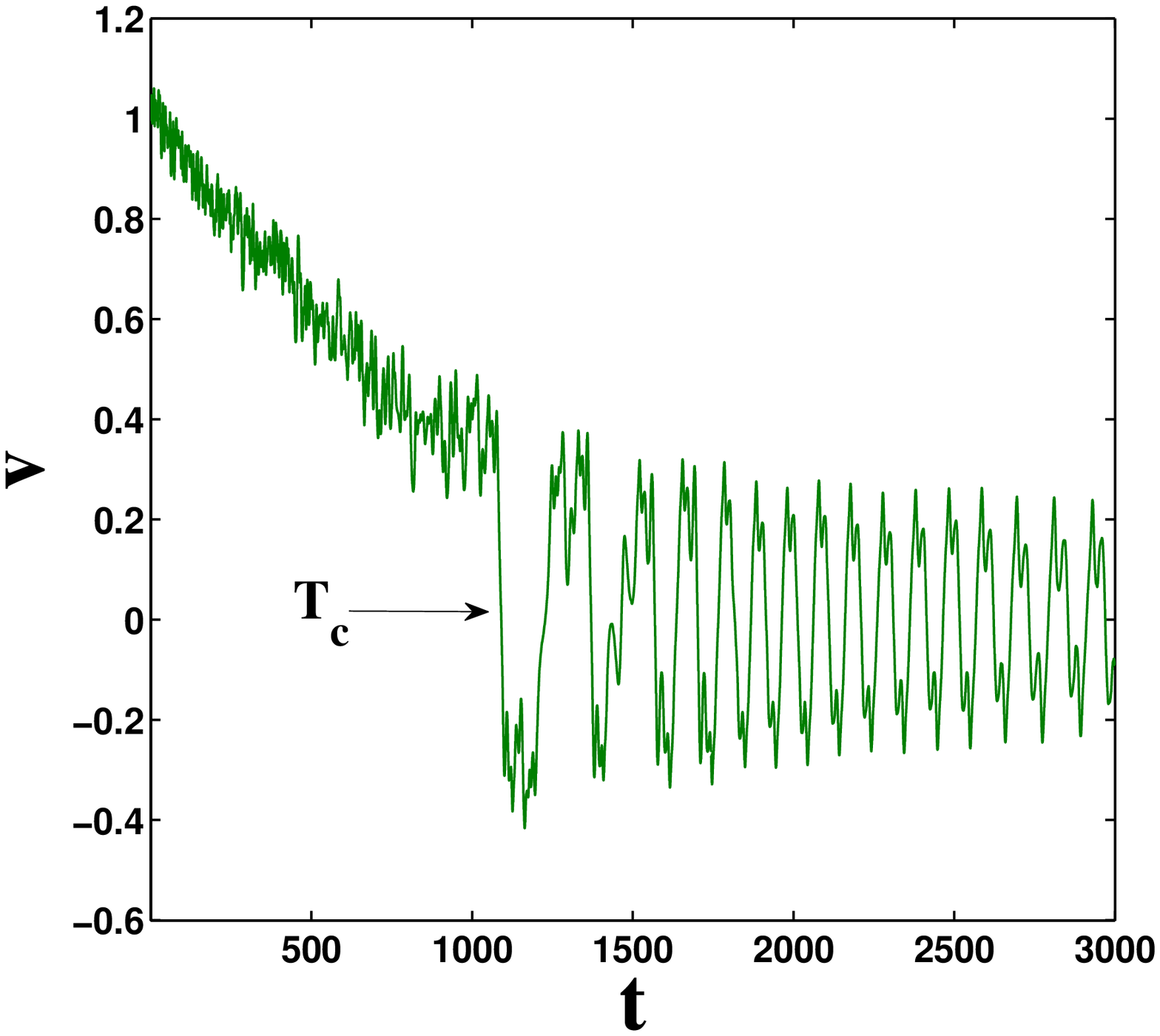}
\vspace{0cm} {\center\footnotesize\hspace{0cm}\textbf{Figure
2(a)}\hspace{6.6cm}\textbf{Figure 2(b)}}\\
\vspace{14mm}
\includegraphics[scale=0.42]{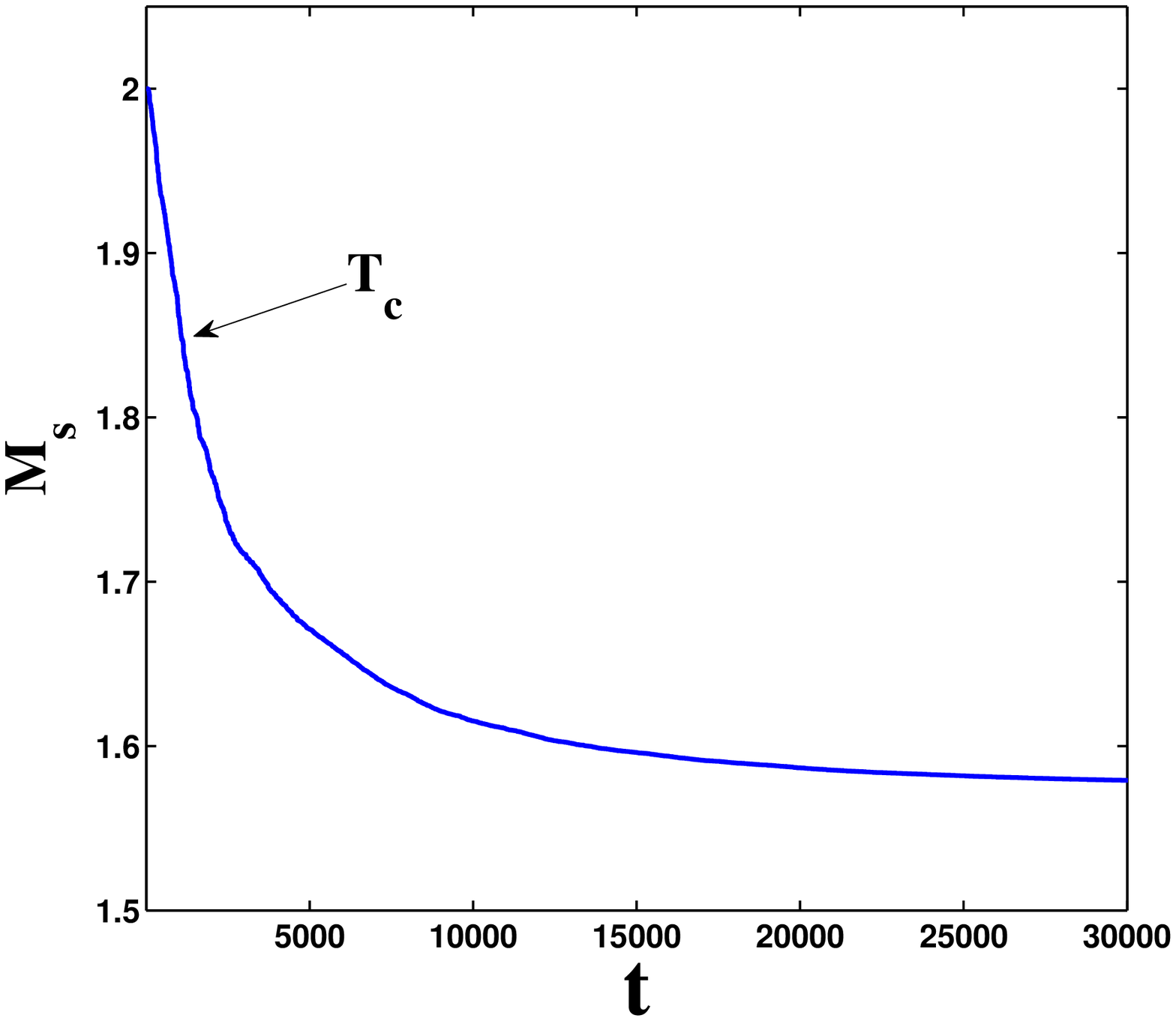}
\vspace{0cm} {\center\footnotesize\hspace{0cm}\textbf{Figure
2(c)}
} \figcaption{AL soliton trapping by one realization of random
potential [$\mu=1$, $v(t=0)=1$, and $W=0.1$]. (a) Center of mass $x$
[Eq.~(\ref{4.5})] as function of time $t$. (b) The soliton velocity
$v$ [Eq.~(\ref{4.7})] as function of time $t$. (c) The mass $M_s$
[Eq.~(\ref{4.4})] as function of time $t$.
} \label{B}
\end{minipage}
\\[\intextsep]
\\[\intextsep]
\begin{minipage}{\textwidth}
\renewcommand{\figurename}{FIGURES }
\renewcommand{\captionfont}{ }
\renewcommand{\captionlabelfont}{ }
\vspace{0cm}\centering
\includegraphics[scale=0.49]{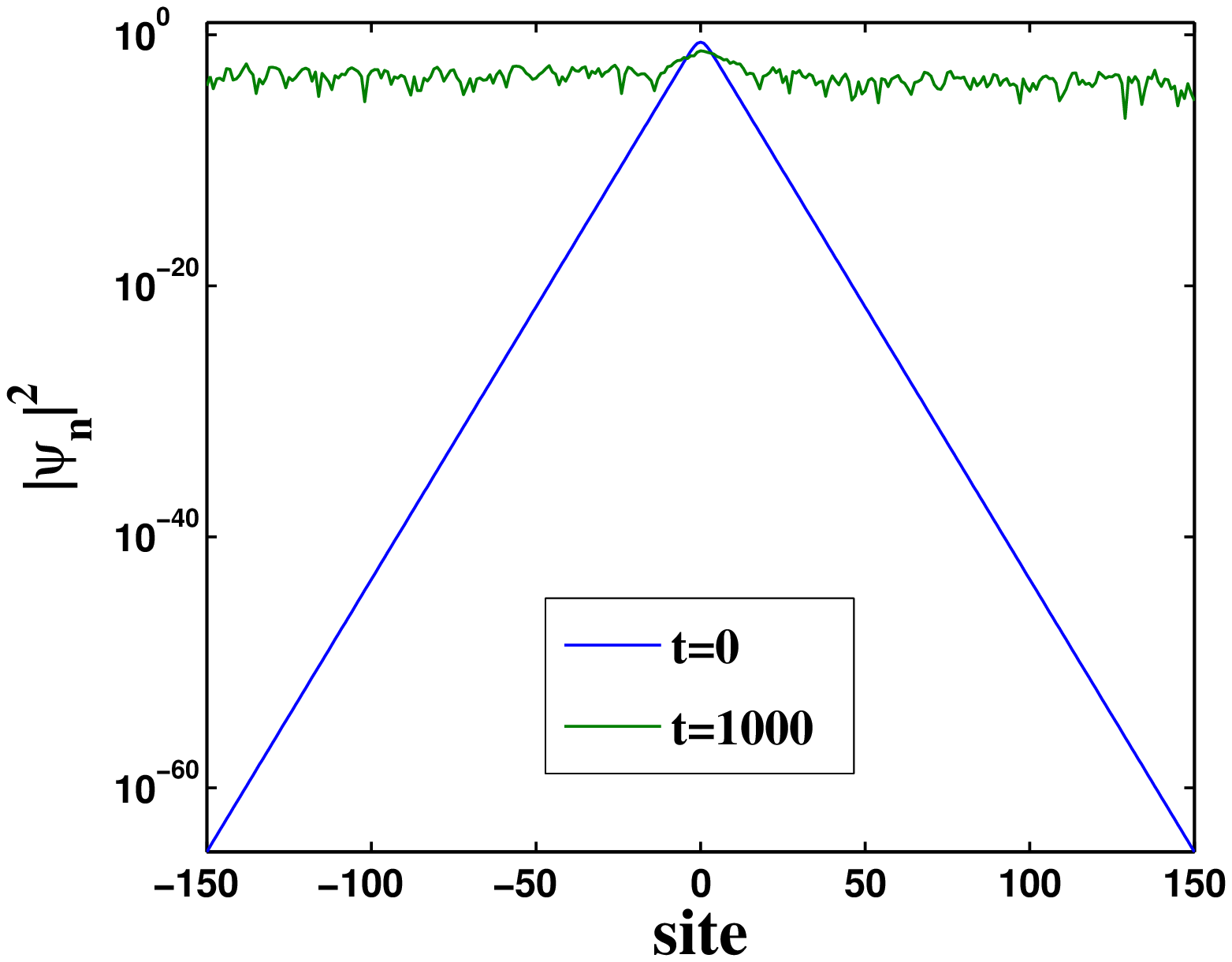}
\includegraphics[scale=0.49]{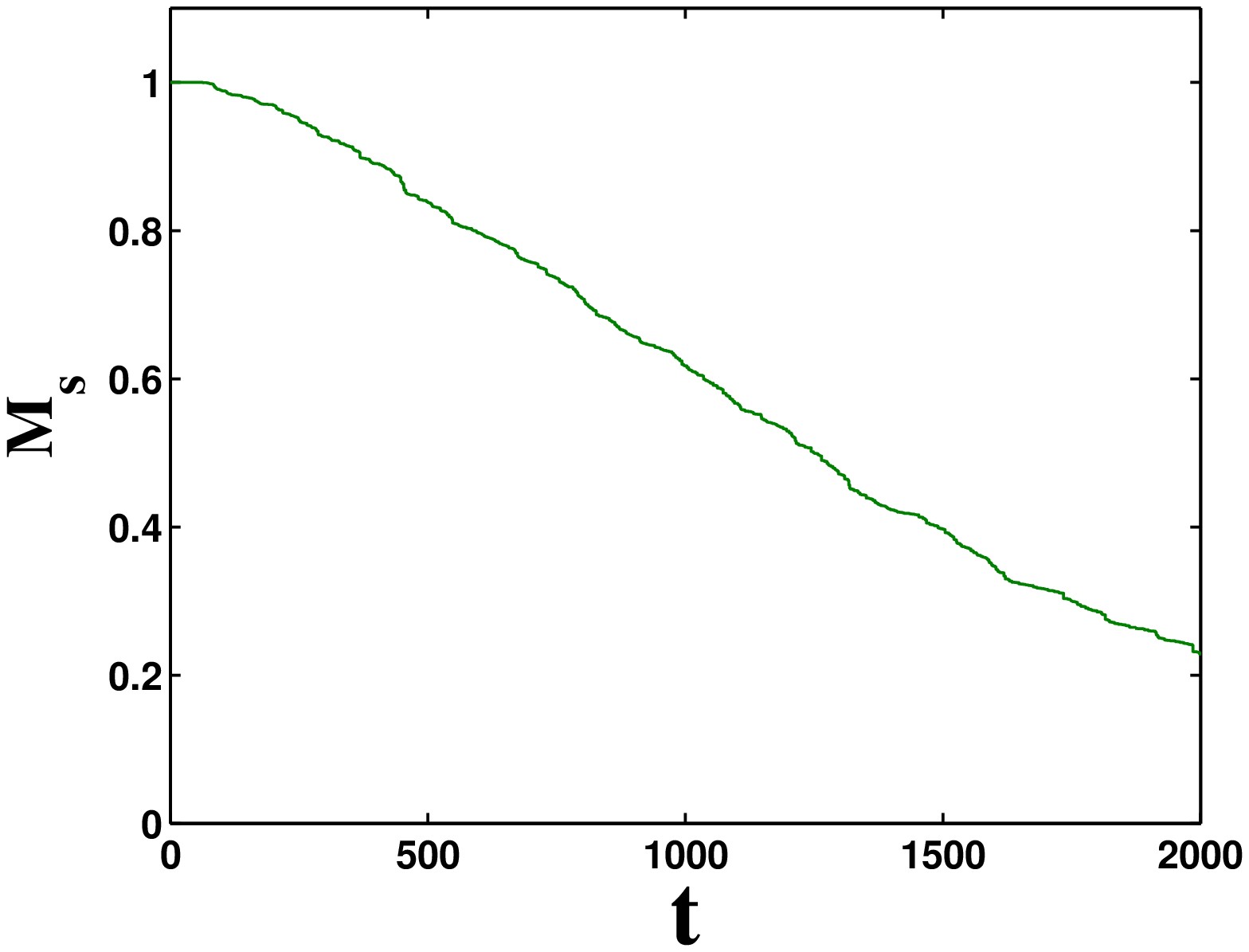}
\vspace{0cm} {\center\footnotesize\hspace{0cm}\textbf{Figure
3(a)}\hspace{6.6cm}\textbf{Figure 3(b)}}\\
\vspace{14mm}
\includegraphics[scale=0.63]{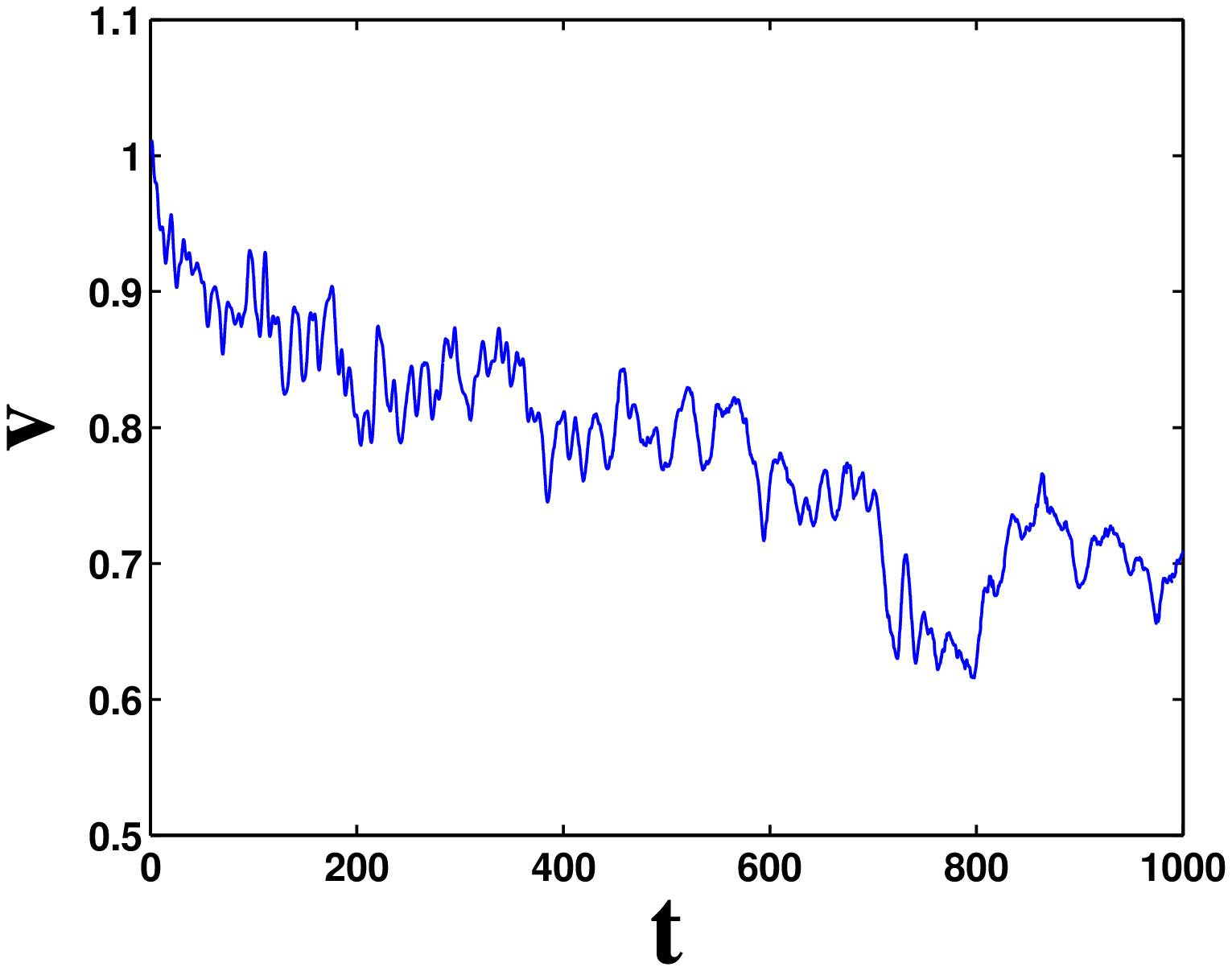}
\vspace{0cm} {\center\footnotesize\hspace{0cm}\textbf{Figure 3(c)}}
\figcaption{AL soliton propagation in one realization of random
potential [$\mu=0.5$, $v(t=0)=1$, and $W=0.1$]. (a) The soliton
profiles at $t=0$ and $t=1000$ (note that the site coordinate is
fixed on the center of mass). (b) The mass $M_s$ [Eq.~(\ref{4.4})]
as function of time $t$. (c) The soliton velocity $v$
[Eq.~(\ref{4.7})] as function of time $t$. } \label{C}
\end{minipage}
\\[\intextsep]
\\[\intextsep]
\begin{minipage}{\textwidth}
\renewcommand{\figurename}{FIGURES }
\renewcommand{\captionfont}{ }
\renewcommand{\captionlabelfont}{ }
\vspace{-1.4cm}\centering
\includegraphics[scale=0.52]{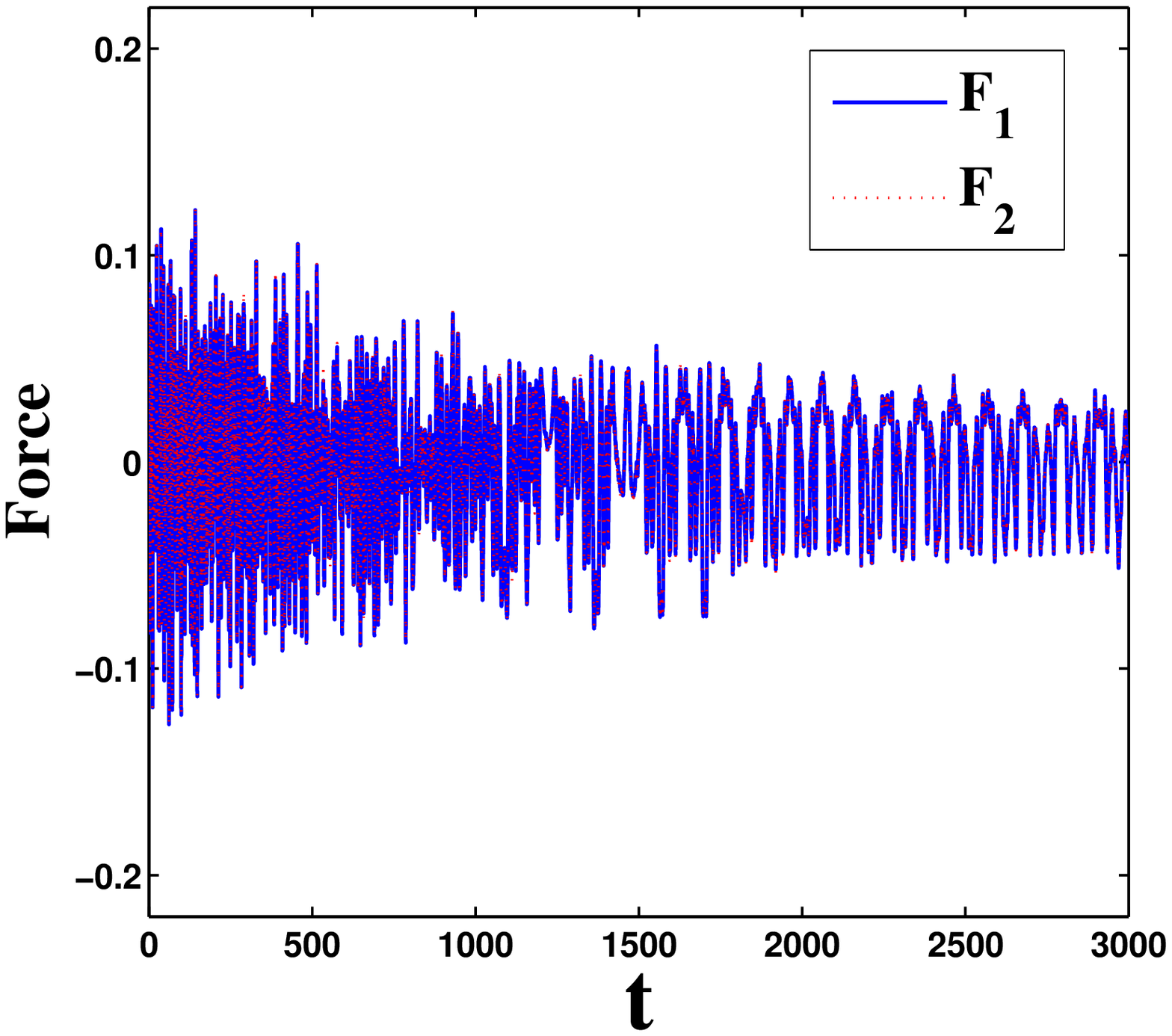}
\vspace{-0.4cm} {\center\footnotesize\hspace{0cm}\textbf{Figure 4(a)}}\\
\vspace{0.5cm}\centering
\includegraphics[scale=0.3]{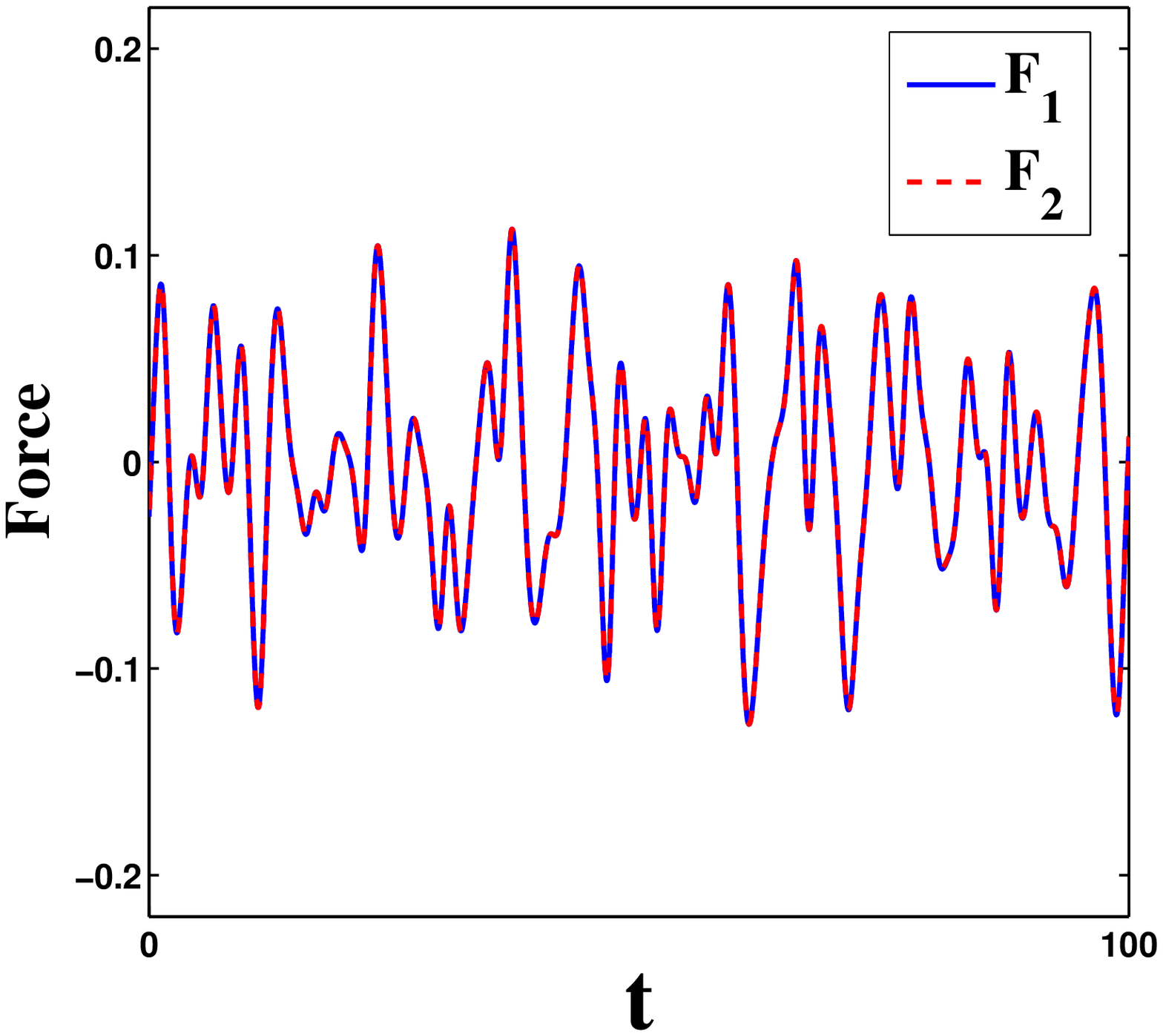}
\includegraphics[scale=0.3]{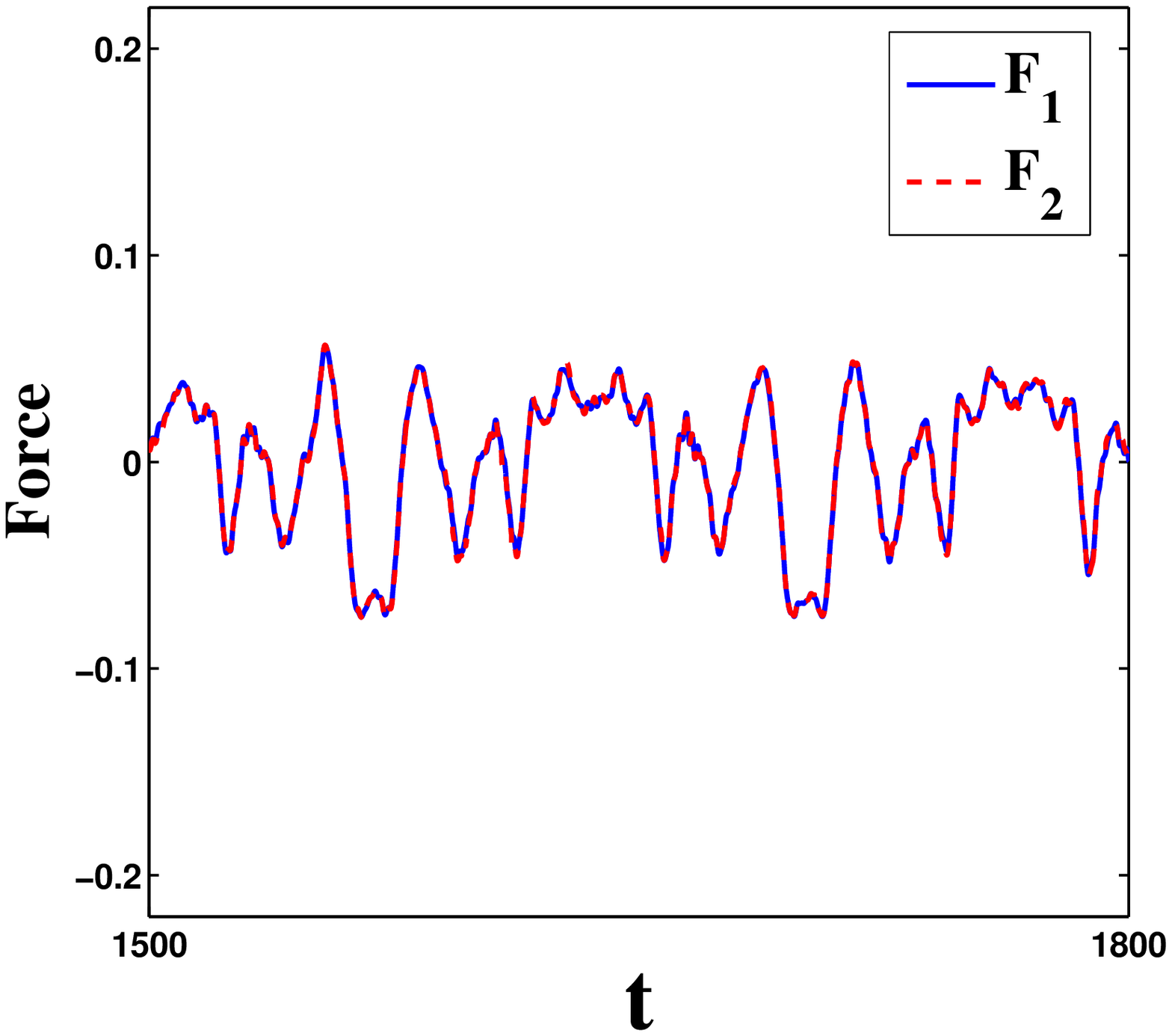}
\includegraphics[scale=0.3]{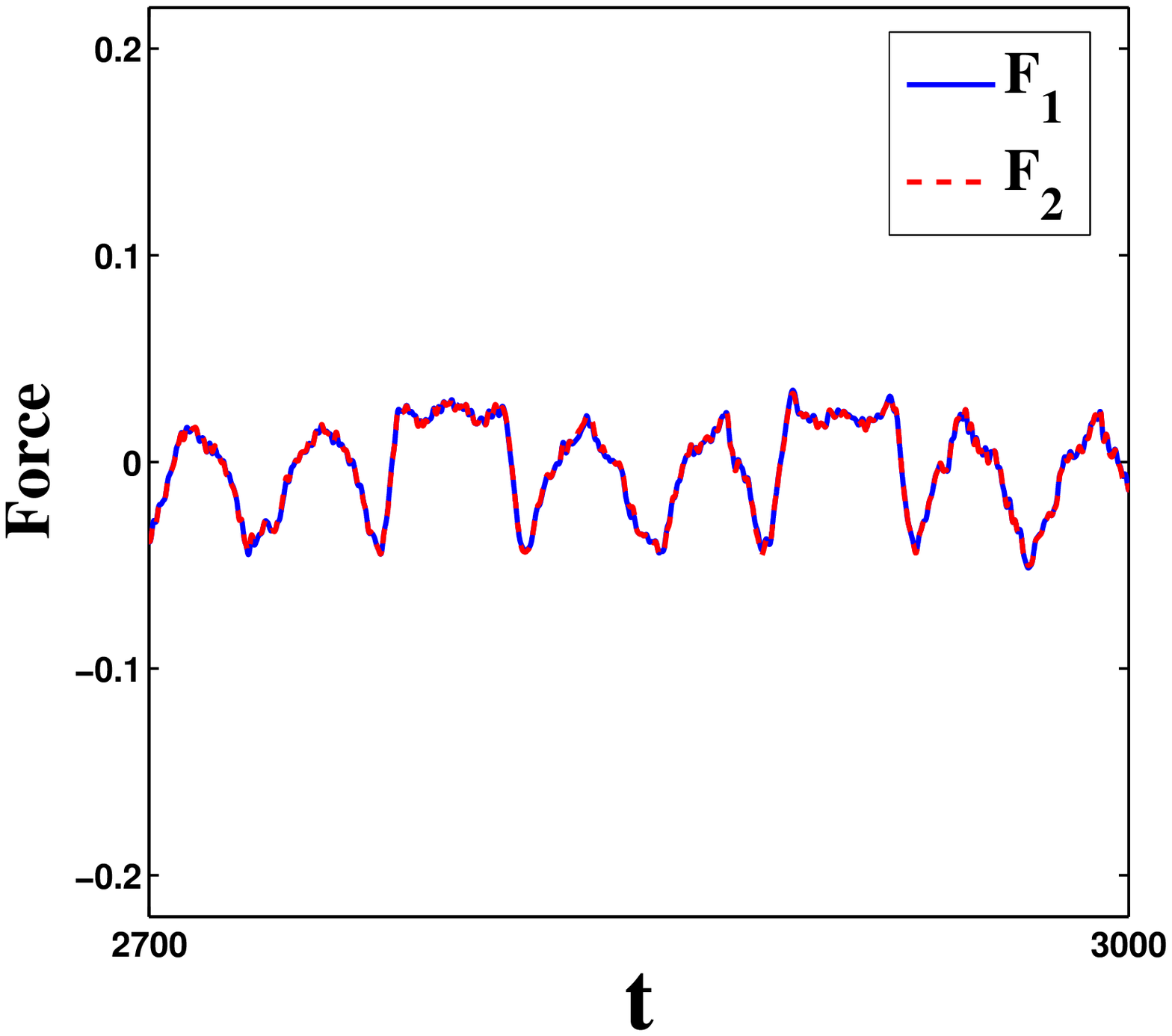}
\vspace{-0.4cm} {\center\footnotesize\hspace{0cm}\textbf{Figure
4(b)}\hspace{3.6cm}\textbf{Figure 4(c)}\hspace{3.6cm}\textbf{Figure 4(d)}}\\
\vspace{0.5cm}\centering
\includegraphics[scale=0.65]{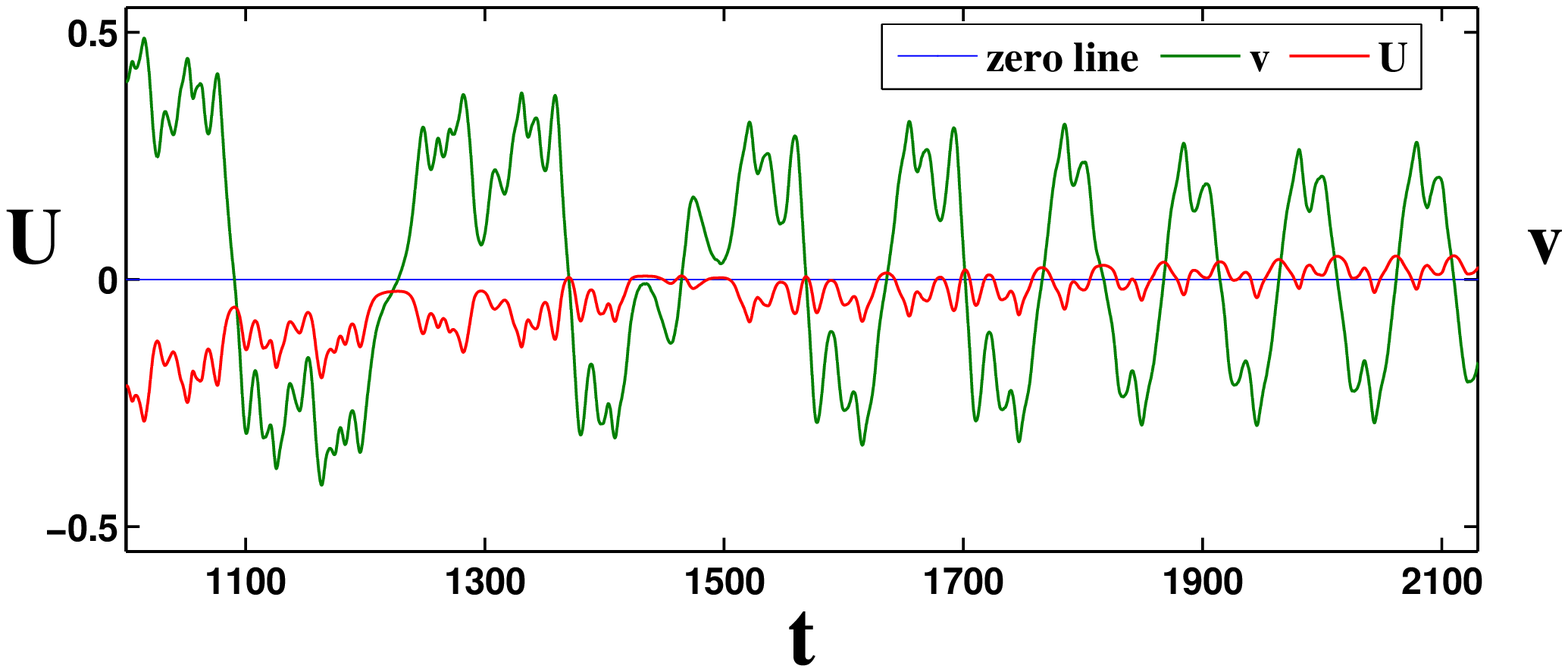}
\vspace{-0.4cm} {\center\footnotesize\hspace{0cm}\textbf{Figure
4(e)}} \figcaption{ (a) Comparison of the two forces $F_{1}$
[Eq.~(\ref{6.1})] and $F_2$ [Eq.~(\ref{6.2})] with the soliton
parameters and random potential the same as those in
Figs.~\ref{B}(a) and (b). (b)-(d) Zoomed views of the panel (a) for
three different time intervals. (e) The effective potential $U$ and
soliton velocity $v$ after the soliton's first reflection
($T_c\approx1100$).} \label{D}
\end{minipage}
\\[\intextsep]
\\[\intextsep]
\begin{minipage}{\textwidth}
\renewcommand{\figurename}{FIGURES }
\renewcommand{\captionfont}{ }
\renewcommand{\captionlabelfont}{ }
\vspace{0cm}\centering
\includegraphics[scale=0.47]{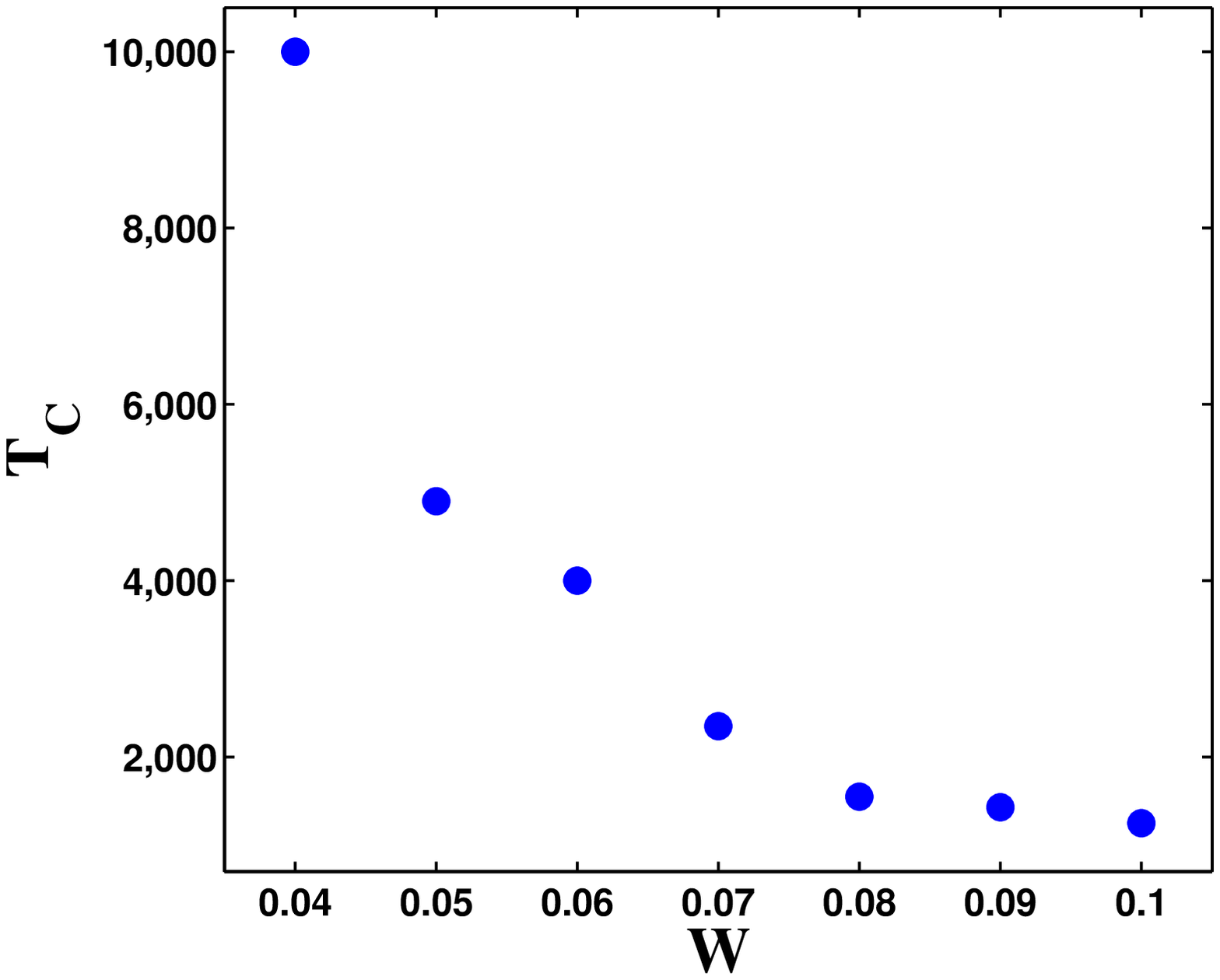}
\includegraphics[scale=0.46]{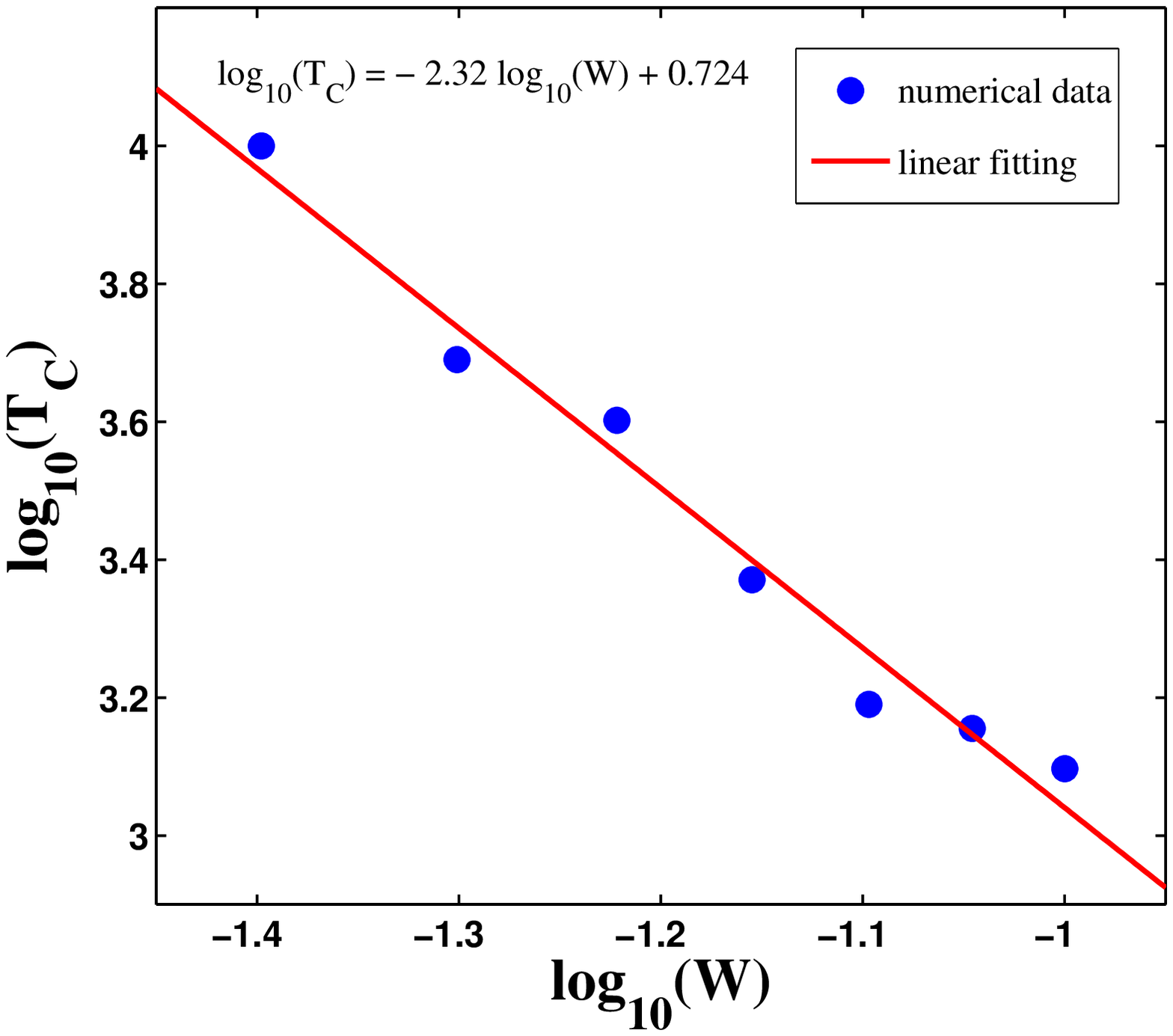}
\vspace{0cm} {\center\footnotesize\hspace{0cm}\textbf{Figure
5(a)}\hspace{6.6cm}\textbf{Figure 5(b)}}\\
\vspace{14mm}
\includegraphics[scale=0.49]{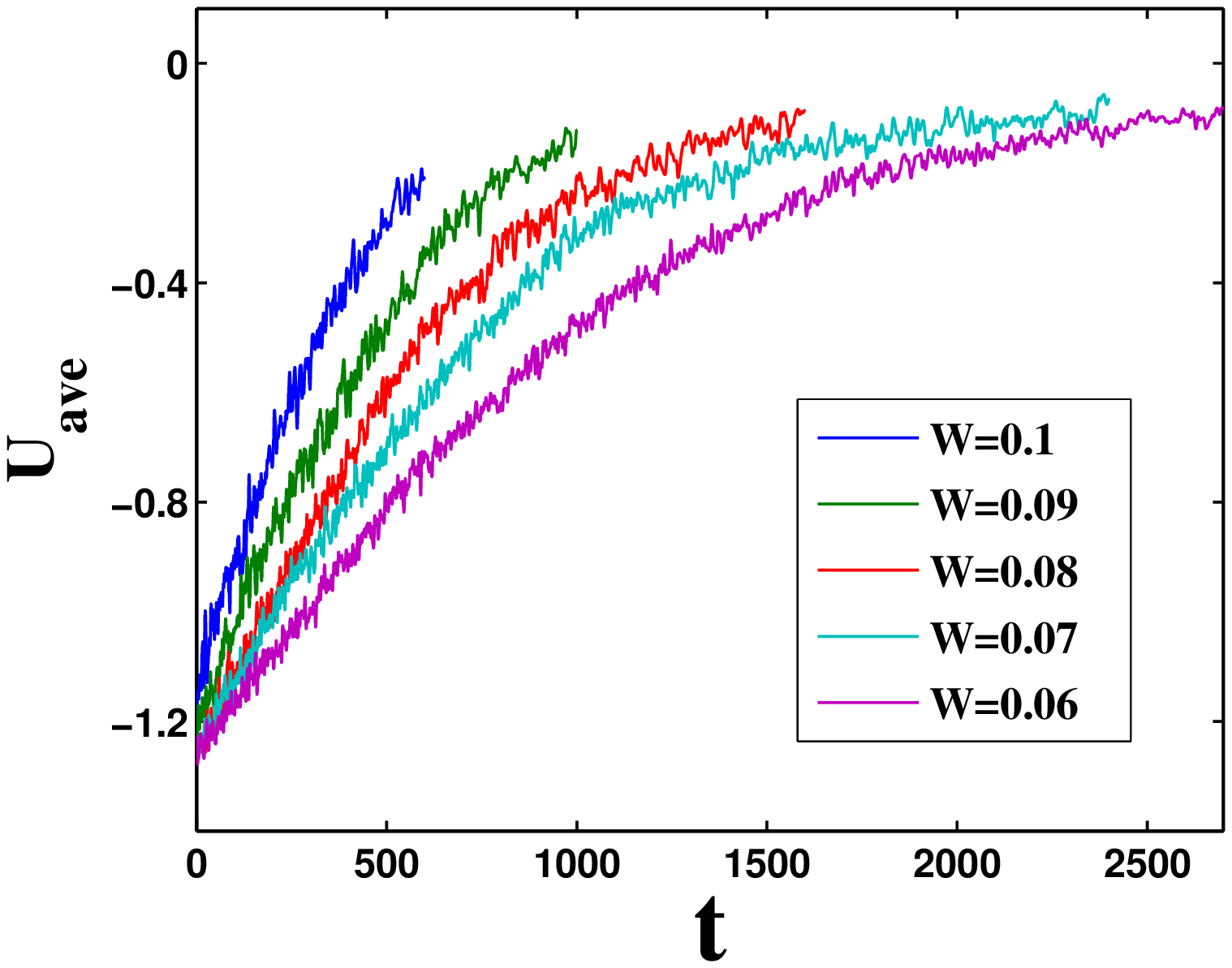}
\includegraphics[scale=0.49]{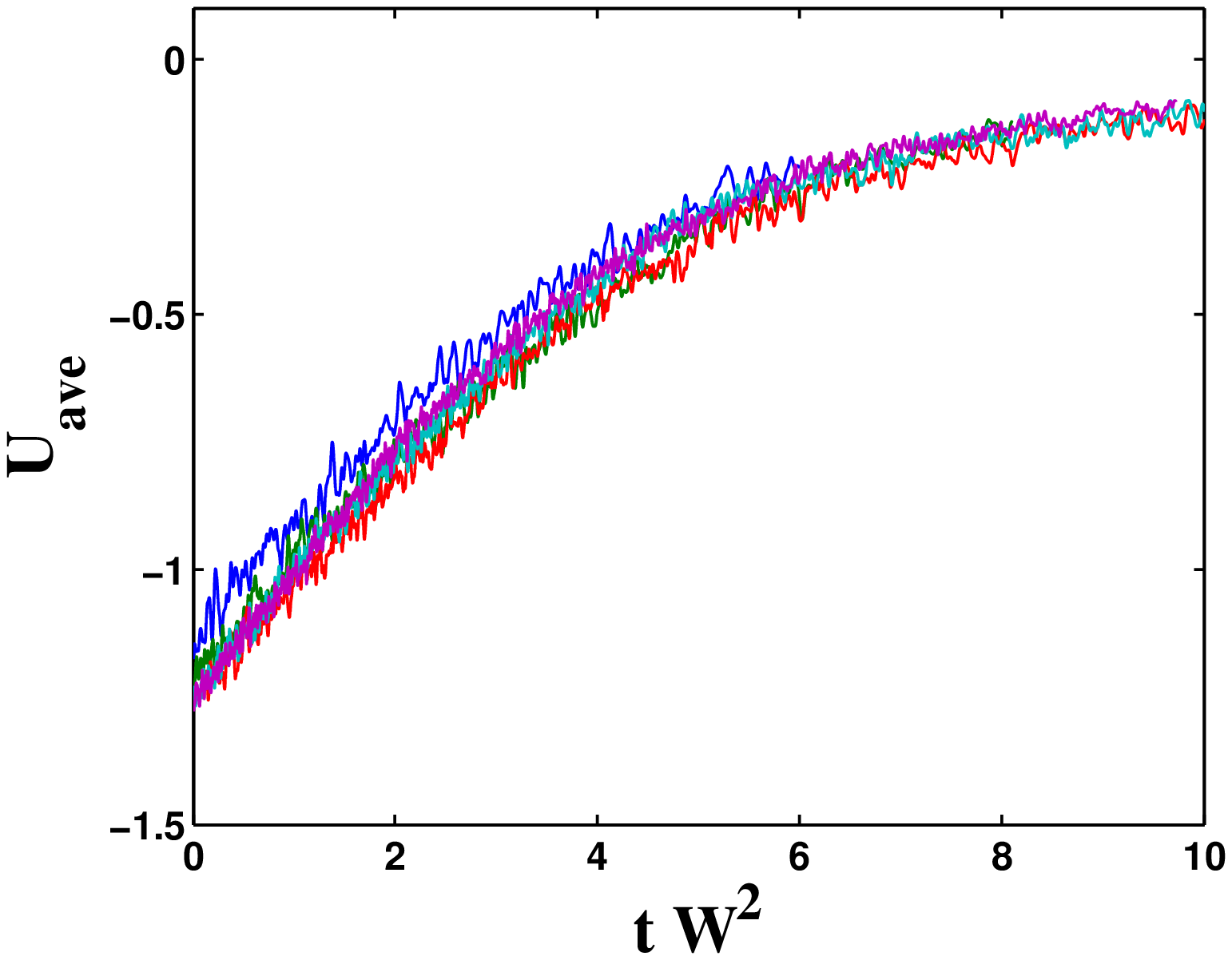}
\vspace{0cm} {\center\footnotesize\hspace{0cm}\textbf{Figure
5(c)}\hspace{6.6cm}\textbf{Figure 5(d)}} \figcaption{ (a) $T_c$ as
function of $W$. The random potential are consisted of one
realization of random numbers in $[-1,1]$, multiplied by different
strength $W/2$. (b) Linear fit of the data (logarithm forms of
variables) in (a). (c) $U_{ave}$ as function of $t$ and $W$. (d)
$U_{ave}$ as a function of the scaling variable $tW^2$. } \label{E}
\end{minipage}
\\[\intextsep]

\end{document}